\documentclass[%
 reprint,
superscriptaddress,
 amsmath,amssymb,
 aps,
prapplied,
]{revtex4-2}
\usepackage{graphicx}
\usepackage{dcolumn}
\usepackage{bm}
\usepackage{hyperref}
\usepackage{todonotes}
\usepackage{float}      

\usepackage{algpseudocode} 
\usepackage{xcolor}

\usepackage{comment}
\usepackage[separate-uncertainty=true]{siunitx}
\usepackage[siunitx]{circuitikz}

\begin{document}

\floatstyle{ruled}      
\newfloat{algorithm}{tbp}{loa}
\floatname{algorithm}{Algorithm }

\preprint{APS/123-QED}

\title{From Cantilevers to Membranes: Advanced Scanning Protocols for Magnetic Resonance Force Microscopy}

\author{Nils Prumbaum}
\affiliation{Laboratory for Solid State Physics, ETH Zürich, CH-8093 Zürich, Switzerland}
\affiliation{Quantum Center, ETH Zürich, CH-8093 Zürich, Switzerland}

\author{Christian L. Degen}
\affiliation{Laboratory for Solid State Physics, ETH Zürich, CH-8093 Zürich, Switzerland}
\affiliation{Quantum Center, ETH Zürich, CH-8093 Zürich, Switzerland}

\author{Alexander Eichler}\thanks{Corresponding author: \href{mailto:eichlera@ethz.ch}{eichlera@ethz.ch}}
\affiliation{Laboratory for Solid State Physics, ETH Zürich, CH-8093 Zürich, Switzerland}
\affiliation{Quantum Center, ETH Zürich, CH-8093 Zürich, Switzerland}

\date{\today}

\begin{abstract}

Magnetic Resonance Force Microscopy (MRFM) enables three-dimensional imaging of nuclear spin densities in nanoscale objects. Based on numerical simulations, we evaluate the performance of strained SiN resonators as force sensors and show that their out-of-plane oscillation direction improves the quality of the reconstructed sample. We further introduce a multislice, compressed-sensing scan protocol that maximizes the information obtained for a given measurement time.
Our simulations predict that these new scanning protocols and optimized algorithms can shorten the total acquisition time by up to two orders of magnitude while maintaining the reconstruction fidelity.
Our results demonstrate that combining advanced scanning protocols with state-of-the-art resonators is a promising path toward high-resolution MRFM for volumetric imaging of biological nanostructures.
\end{abstract}

\maketitle

\section{Introduction}
Magnetic Resonance Force Microscopy (MRFM) is a method for detecting electron or nuclear spins within three-dimensional nanoscale objects~\cite{budakian_roadmap_2024}. Invented in the early 1990s~\cite{sidles_noninductive_1991,sidles_folded_1992}, MRFM achieved many experimental milestones, including the measurement of single electron spins~\cite{rugar_single_2004}, imaging of nuclear spin densities inside a single tobacco mosaic virus~\cite{degen_nanoscale_2009}, Fourier-encoded spin detection~\cite{nichol_nanoscale_2013}, one-dimensional imaging with sub-nanometer resolution~\cite{grob_magnetic_2019,rose_high-resolution_2018}, and dynamic nuclear polarization in nanoscale samples~\cite{tabatabaei_large-enhancement_2024}. To reach these milestones, the sensitivity of MRFM was improved by operating experiments at millikelvin temperatures~\cite{haas_nuclear_2022}, engineering high-gradient magnetic field sources~\cite{longenecker_high-gradient_2012,pachlatko_nanoscale_2024}, and replacing pendulum-style cantilevers~\cite{mamin_silicon_2012} with silicon nanowires~\cite{nichol_nanoscale_2013,sahafi_ultralow_2020}. In spite of all these efforts, MRFM imaging has not reached a level where the structure of biological, chemical, or solid-state samples is revealed at near-atomic scales.

Over the past decade, resonators made from strained silicon nitride opened up many new possibilities in nanomechanical sensing~\cite{engelsen_ultrahigh-quality-factor_2024}. These resonators come in the shape of trampolines~\cite{reinhardt_ultralow-noise_2016,norte_mechanical_2016}, membranes~\cite{tsaturyan_ultracoherent_2017,rossi_measurement-based_2018,reetz_analysis_2019}, strings~\cite{ghadimi_elastic_2018,beccari_strained_2022,gisler_soft-clamped_2022}, polygons~\cite{bereyhi_perimeter_2022}, hierarchical structures~\cite{fedorov_fractal-like_2020,bereyhi_hierarchical_2022}, and spider webs~\cite{shin_spiderweb_2022}. Due to their ultralow dissipation, they have the potential to vastly improve the speed and spatial resolution of MRFM measurements~\cite{eichler_ultra-high-q_2022}. Their high resonance frequencies also avoid issues arising with ultrasoft cantilevers, such as non-contact friction~\cite{heritier_spatial_2021,engelsen_ultrahigh-quality-factor_2024} and static bending~\cite{krass_force-detected_2022}. Recent proof-of-principle experiments demonstrated scanning force microscopy~\cite{halg_membrane-based_2021,gisler_enhancing_2024} and electron spin detection~\cite{scozzaro_magnetic_2016,fischer_spin_2019} with membrane and trampoline resonators.

In spite of these advances, several issues need to be overcome to exploit silicon nitride resonators for nuclear spin detection. The high resonance frequencies of silicon nitride resonators, which typically lie in the range of \SI{100}{\kilo\hertz} to \SI{5}{\mega\hertz}, present a challenge to nuclear spin manipulation. Several solutions have been proposed, including frequency-modulation MRFM~\cite{kuehn_advances_2008}, time-dependent gradients~\cite{nichol_nanomechanical_2012}, spin-induced parametric coupling between nearby modes~\cite{kosata_spin_2020}, and resonant coupling between spins and resonator~\cite{visani_near-resonant_2025}. Moreover, the arrangement of a membrane- or string-based setup is different from that of a pendulum-style cantilever or nanowire-based setup, leading to different point spread function (PSF). This is potentially important, as the PSF determines the spatial convolution of the sample object into measurement data, and consequently impacts the quality of the image reconstruction.

In this work, we present advanced techniques for three-dimensional image reconstruction in MRFM, applicable to setups employing strained silicon nitride-based resonators. Using numerical simulations, we show that a two-dimensional spatial scan combined with a frequency dimension improves the reconstruction fidelity by 2-5x compared to the traditional three-dimensional spatial scan. The improvement is more pronounced for membrane-type sensors, oscillating vertically, compared to cantilever-type sensors oscillating laterally. We also show that compressed sensing allows for a sub-sampling by at least 50\%, permitting a speed-up by the same factor without degradation of the image. Our work indicates avenues for improved three-dimensional MRFM that take advantage of new generations of ultrasensitive nanomechanical resonators.

The manuscript is structured as follows: in Section~\ref{sec:stud_geo}, we compare the different geometries of setups utilizing a cantilever or a membrane resonator as a sensor. In Section~\ref{sec:image_protocols}, we discuss various signal generation and imaging protocols. In Section~\ref{sec:reconstruction}, we explain how the simulated data is reconstructed into a spin distribution image. Finally, in Section~\ref{sec:res}, we show the reconstructed spin density images and analyze the performance of different geometries as well as different imaging protocols.

\section{Setup Geometries} \label{sec:stud_geo}

\begin{figure*}[!htb]
	\centering
	\includegraphics[width=\textwidth]{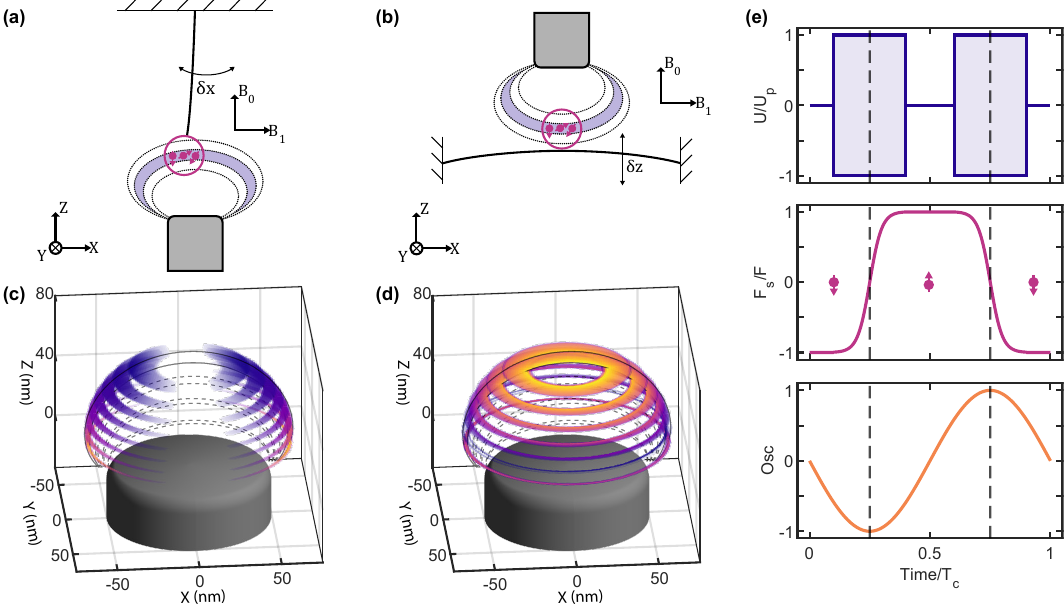}
	\caption{Analyzed geometries. (a)~Illustration of MRFM setup utilizing cantilever-style force sensors. The spin ensemble (pink) is attached to the tip of the resonator and positioned above the magnetic field gradient source (gray). The resonator's oscillation direction is along $x$ and parallel to the gradient source's top surface. When spin inversion pulses are applied via an external antenna, all spins inside the resonance slice (purple) are inverted periodically. (b)~Illustration of a setup utilizing tensioned resonators as force sensors. The sample is still attached to the resonator, but the oscillation direction is along $z$, perpendicular to the gradient source's top surface, changing the direction of the detected force. (c)~PSF of a cylindrical nanomagnet when the cantilever geometry is used. Its magnitude is proportional to $|\partial B_z/\partial x|^2$.  The nanomagnet's radius is $\SI{50}{\nano\metre}$ and it has a rounded edge with radius $r_\mathrm{edge} = \SI{10}{\nano\metre}$. (d)~PSF of the identical magnet for a membrane resonator geometry. The strength of the PSF is determined by the derivative $|\partial B_z/\partial z|^2$. (e)~Timing diagram of spin inversions. Twice per resonator oscillation, a spin inversion pulse (blue) is applied. This results in a force generated by the ensemble (pink) that is resonant with the resonator's movement (orange) leading to a detectable driving force.}
	\label{Fig:1} 
\end{figure*}

\textit{Cantilever-based setup.} Typical MRFM setups utilize a pendulum-style cantilever as a sensor~\cite{degen_nanoscale_2009}, see Fig.~\ref{Fig:1}(a). The cantilever oscillates along the $x$-axis, parallel to the surface of a nanoscale ferromagnet that acts as a magnetic field gradient source. Spins in a sample attached to the cantilever tip experience a force $\mathbf{F}_\mathrm{{spin}}(t) = \nabla  (\mathbf{m}(t) \cdot \mathbf{B})$, where $\mathbf{B}$ is the local magnetic field and $\mathbf{m}(t)$ the time-dependent magnetic moment of the spin ensemble. The spin quantization axis is set by a strong external magnetic field $B_0$ in the $z$-direction. The $z$-component of the magnetic moment, $m_z(t)$, is periodically inverted using radio-frequency (rf) pulses resonant with the Larmor frequency of the spins. The rf magnetic field, of magnitude $B_1$, points along the $x$-direction and is generated by a microstrip antenna~\cite{poggio_nuclear_2007}. Since the resonator is only susceptible to forces along its oscillation direction, here the $x$-direction, the effective driving force is 
\begin{equation}
    F(t) = \frac{\partial B_z}{\partial x}m_z(t).
    \label{eq:ForceCant}
\end{equation}
Note that interchanging the positions of the sample and the magnetic field gradient source~\cite{kuehn_advances_2008,vinante_magnetic_2011,issac_dynamic_2016} does not change Eq.~\ref{eq:ForceCant}.


\textit{Membrane-based setup.} Strained silicon nitride mechanical resonators have been fabricated in many different geometries, such as membranes, strings or trampolines~\cite{reinhardt_ultralow-noise_2016,norte_mechanical_2016,tsaturyan_ultracoherent_2017,rossi_measurement-based_2018,reetz_analysis_2019,ghadimi_elastic_2018,beccari_strained_2022,gisler_soft-clamped_2022,bereyhi_perimeter_2022,fedorov_fractal-like_2020,bereyhi_hierarchical_2022,shin_spiderweb_2022}. In this work, we use membranes as a generic example of all these resonator types. In Fig.~\ref{Fig:1}(b), we show a sketch of a setup with a membrane sensor~\cite{halg_membrane-based_2021}. Here, the sample containing spin ensembles is placed on the membrane surface. Importantly, in contrast to pendulum-style cantilevers, the membrane's oscillation is along the $z$ direction. This geometry leads to a driving force in the $z$-direction and given by
\begin{equation}
    F(t) = \frac{\partial B_z}{\partial z} m_z(t).
    \label{eq:ForceMem}
\end{equation}

\section{Imaging Protocols}\label{sec:image_protocols}
\subsection{Single Point}\label{sec:single_point}
First, we examine how a resonant force signal is generated for a single magnetic moment located at a fixed position with respect to the magnetic field gradient. We start with the simple case of a point-like magnetic moment $\mu_z$. This produces a static force $F_\mathrm{res}=\frac{\partial B_z}{\partial p}\mu_z$, where $p \in \left[x,z\right]$ is the oscillation direction of the resonator. To generate a resonant force, rf inversion pulses are applied, cf. blue trace in Fig.~\ref{Fig:1}(e)~\cite{degen_nanoscale_2009,grob_magnetic_2019}. These pulses flip the $z$-component of the spin's magnetic moment and thereby periodically reverse the sign of the force [Eq.~\ref{eq:ForceCant}]. By setting the inversion pulse repetition rate to twice the natural frequency of the resonator, $1/T_\mathrm{r} = 2f_\mathrm{res}$, the induced force is brought into resonance with the mechanical resonator, see pink and orange traces in Fig.~\ref{Fig:1}(e).

In the next step, we consider an extended spin ensemble, characterized by a local spin density $\rho_\mathrm{spin} (\mathbf{r})$. Spins located at different positions relative to the gradient source possess different Larmor frequencies $f_\mathrm{L}(\mathbf{r}) = \gamma/2\pi~|\mathbf{B}(\mathbf{r})|$ depending on the local magnetic field strength $|\mathbf{B}(\mathbf{r})|$ and the gyromagnetic ratio $\gamma$ of the detected spin species. An inversion pulse is characterized by a center frequency $f_0$ and an inversion bandwidth $f_\mathrm{BW}$, and only those spins whose Larmor frequencies lie within the inversion range $f_0\pm f_{\mathrm{BW}}$ of the applied pulse are inverted. For a given pulse, this creates a `resonant inversion slice', shown in purple in Figure \ref{Fig:1}(a-b)~\cite{degen_nanoscale_2009}.

In nanoscale spin ensembles, the thermal (Boltzmann) spin polarization is typically negligible. Instead, stochastic fluctuations of the spin ensemble polarization dominate~\cite{degen_role_2007,herzog_boundary_2014}. Therefore, it becomes advantageous to analyze the force variance, rather than its mean, for spin imaging. The net resonant force variance is generated by the sum of all spins inside this resonance slice, weighted by their respective local gradients.
This spatial distribution is represented in a `point-spread function' (PSF), which we refer to as $H(\mathbf{r})$, where $\mathbf{r} = (x,y,z)$ is the spatial position with respect to the magnetic gradient source. Neglecting pulse imperfections, the PSF for a given inversion pulse and magnetic field gradient source is expressed as 
\begin{equation}
    H(\mathbf{r}) = 
  \begin{cases}
     \sigma^2_\mathrm{spin} &  \frac{\gamma}{2\pi} B(\mathbf{r})\in (f_0 \pm f_\mathrm{BW})\\
    0 & \text{otherwise}.
  \end{cases}
\end{equation}
where $\sigma^2_\mathrm{spin} = \left(\frac{\partial B_z(\mathbf{r})}{\partial_x}\right)^2\mu^2$ (for cantilevers) or $\sigma^2_\mathrm{spin} = \left(\frac{\partial B_z(\mathbf{r})}{\partial_z}\right)^2\mu^2$ (for membranes) is the force variance generated by a single moment $\mu = |m_z| = \frac{1}{2}\hbar \gamma$ at position $\mathbf{r}$, see Eq.~\ref{eq:ForceCant} and \ref{eq:ForceMem}. 
We compute the nanomagnet's magnetic field $B(\mathbf{r})$ using the FEMM finite-element solver~\cite{meeker_finite_2023}.

Figures~\ref{Fig:1}(c-d) show examples of PSFs for the two geometries discussed in Section~\ref{sec:stud_geo}, both employing the same cylindrical nanomagnet with rounded edges. For the cantilever, where forces are strongest at the magnet's  tangential point to the $y$ direction, the PSF has its maximum contribution in this region near the magnet's surface and decays rapidly with increasing distance [Fig.~\ref{Fig:1}(c)]. 
In contrast, the membrane is susceptible to forces $\propto\partial B_z/\partial z$, and the strongest gradients inside the resonance slice are located over the center of the magnet [Fig.~\ref{Fig:1}(d)]. Although the maximum gradient of the membrane-based setup is smaller compared to the cantilever-based setup, it decreases more slowly with distance from the surface, thereby enhancing the signal contribution of distant parts of the sample.

The total force variance detected by the sensor at a fixed position of the resonator $\mathbf{r}_i = (x_i,y_i,z_i)$ with respect to the magnetic field gradient source is given by the three-dimensional sum over all magnetic moments
 \begin{equation}
     \sigma^2_\mathrm{tot}(\mathbf{r}_i) = \sum_{\mathbf{r}} H(\mathbf{r}) \rho_\mathrm{spin}(\mathbf{r}-\mathbf{r}_i) + \sigma_\mathrm{noise}^2,
 \label{eq:spin_reponse}
 \end{equation}
where $\sigma_\mathrm{noise}^2$ represents additive measurement noise consisting of two primary sources. The first source arises from the thermomechanical force noise of the mechanical resonator, which can be modeled as a stochastic force with variance $\sigma_\mathrm{th}^2 = 4k_\mathrm{B} T m \Gamma \Delta f$. Here,  $k_\mathrm{B}$ is the Boltzmann constant, $T$ is the resonator mode temperature, $m$ the mode mass, and $\Delta f$ the measurement bandwidth. The dissipation coefficient $\Gamma = \omega_0/Q$ (with $Q$ the quality factor and $\omega_\mathrm{0}$ the angular resonance frequency) is a key figure of merit for mechanical resonators, describing the ratio of the resonator's total energy to the energy lost per oscillation. Together with the mass, it is a primary optimization parameter in the development of improved resonator geometries to minimize thermomechanical force noise.

The second source of uncertainty stems from the fact that the detected quantity is an estimated force variance. Within a finite measurement time $T_\mathrm{m}$, this estimator has an intrinsic uncertainty, governed by the spin noise correlation time $\tau_\mathrm{m}$~\cite{degen_role_2007,slichter_principles_1996}. The resulting total measurement uncertainty is~\cite{degen_role_2007}
\begin{equation}
    \sigma_\mathrm{noise}^2 \approx \left(\frac{2}{\frac{T_\mathrm{m}}{\tau_\mathrm{m}}-1}\right)^{1/2}\left(\sigma_\mathrm{spin}^4+2\sigma_\mathrm{th}^2\sigma_\mathrm{spin}^2+2\sigma_\mathrm{th}^4\right)^{1/2},
    \label{eqn:noise}
\end{equation}
where $T_\mathrm{m}/\tau_\mathrm{m}$ is the number of independent spin configurations sampled per measurement.\\
To account for the discrete sampling inherent to MRFM scanning experiments and to simplify the subsequent description of algorithms, we adopt a matrix notation. We describe the measurement volume using a regular grid of points $\mathbf{r}_{i,j,k} = (x_i,y_j,z_k)$, where $i\in[1,n_x],j\in[1,n_y],k\in[1,n_z]$. Here, $n_x,n_y,n_z$ are the number of grid points along the $x$, $y$, and $z$ direction, respectively. Evaluating $H(\mathbf{r})$ at every grid point results in the three-dimensional matrix $\mathbf{H}$ of size $(n_x,n_y,n_z)$ with entries $ H_{i,j,k} = H(\mathbf{r}_{i,j,k})$. In the same way, matrices are defined for the sample's local spin density $\mathbf{O}$ with entries $ O_{i,j,k} = \rho_\mathrm{spin}(\mathbf{r}_{i,j,k})$ and the measured force variance $\mathbf{I}$ with $I_{i,j,k} = \sigma_\mathrm{tot}^2(\mathbf{r}_{i,j,k}).$

\subsection{XYZ Scan}
In a conventional MRFM experiment, the magnetic field gradient source is scanned relative to the sample on a regularly spaced three-dimensional grid~\cite{degen_nanoscale_2009,krass_force-detected_2022}, which we refer to as an `XYZ scan'. At each point on this grid, a single measurement of duration $T_\mathrm{m}$ is performed, during which the same rf inversion pulse (fixed $f_0$ and $f_{\rm BW}$), that defines a single resonant slice, is applied, cf. Sec.~\ref{sec:single_point}. The resulting scan is described by
\begin{equation}
\label{eq:xyz}
    \mathbf{I}_\mathrm{XYZ} = \mathbf{H} \star_{(x,y,z)} \mathbf{O}+ \bm\eta,
\end{equation}
where $\mathbf{I}_\mathrm{XYZ}$ denotes the recorded measurement data, $\star_{(x,y,z)}$ is the three-dimensional convolution operator and $\bm\eta$ with $\eta_{i,j,k} = \sigma^2_\mathrm{noise}(r_{i,j,k})$, is voxel-wise additive noise as described in Eq.~\eqref{eqn:noise}.\\
A key design consideration for such an MRFM experiment is the selection of the spin inversion pulse parameters. The inversion bandwidth $f_\mathrm{BW}$ determines the width of the resonant slice, while the center frequency $f_0$ influences the maximum distance from the nanomagnet's surface at which spin inversions occur. On the one hand, $f_0$ must be chosen such that the PSF extends at least beyond the sum of the sample diameter and the minimum standoff distance. On the other hand, if the PSF extends too far, the magnetic field gradients become very small, reducing the signal contribution from individual spins. This imposes a trade-off between scan depth and sensitivity.

\subsection{Multislice Scan}
In the following, we propose an alternative scanning protocol, termed the `multislice scan', that better addresses the trade-off between the scan depth and sensitivity. In this protocol, we scan the gradient source only in a single $(x,y)$ plane, rather than a series of planes at different $z$-heights. At each grid position within this plane, $n_\mathrm{f}$ subsequent measurements are made, each utilizing a spin inversion pulse with a unique combination of center frequency $f_{0,\mathrm{l}}$ and inversion bandwidth $f_{\mathrm{BW,l}}$. This produces $n_\mathrm{f}$ different data points according to Eq.~\ref{eq:spin_reponse} for each position $(x_i,y_j)$ on the $(x,y)$ plane, corresponding to the response of the sample to the $n_\mathrm{f}$ PSFs of the spin inversion pulses. Previously, MRFM scanning protocols utilized multiple inversion pulses~\cite{zuger_threedimensional_1996,wago_paramagnetic_1998,tsuji_three_2006} in a geometry where different pulse frequencies can be approximated as a shift of the resonance slice in the $z$-direction. In contrast, our geometry leads to different PSF shapes for different pulse frequencies. This makes it necessary that we recompute the PSF for the parameters of each applied inversion pulse.

The multislice scan is described in matrix notation by

\begin{equation}
    \mathbf{I}_{\mathrm{MS},l} = \sum_{z = 1}^{n_z} \mathbf{H}^l\star_{(x,y)}\mathbf{O} + \bm\eta, 
    \label{eq:multislice}
\end{equation}
where $\mathbf{I}_\mathrm{MS}$ is the measurement matrix of size $(n_x,n_y,n_f)$ with $\mathbf{I}_{\mathrm{MS},l}$ being the $l$-th $(x,y)$ plane of this matrix, $\mathbf{H}^l$ is the matrix corresponding to the PSF of the $l$-th inversion pulse parameter set, and $\star_{(x,y)}$ denotes the slice-wise two-dimensional convolution along the $(x,y)$-plane. The individual entries of the measurement matrix $\mathbf{I}_\mathrm{MS}$ are given by 
\begin{equation}
    I_{\mathrm{MS},i,j,l} = \sum_{i^{'}=1}^{n_x} \sum _{j^{'}=1}^{n_y}\sum_{k=1}^{n_z} H^l_{i-{i^{'}},j -{j^{'}},k}O_{{i^{'}},{j^{'}},k} + \sigma^2_\mathrm{noise}
\end{equation}

Compared to the XYZ scan using a fixed resonance slice, the multislice scanning protocol offers significant advantages. Most notably, it overcomes the trade-off between extending the PSF far into the $z$ direction and achieving high signal contributions from individual spins. In the multislice method, for each height $z$, we can use the rf pulse parameters that generate the optimal resonant slice, which is the one containing the highest gradient for measuring spins at that particular height.
In addition, the measurement efficiency is improved by reducing the number of different scanning positions, thereby reducing the need for resonator recalibration steps due to interactions with the nanomagnet surface~\cite{krass_force-detected_2022,pachlatko_nanoscale_2024,heritier_spatial_2021}. Finally, keeping the resonator stationary during the $n_\mathrm{f}$ measurements allows for a prolonged estimation of the resonator's thermomechanical force noise. The resulting total measurement uncertainty, as derived in Appendix~\ref{app:nosie_est}, is given by
\begin{equation}
\label{eq:noisepoint}
\begin{split}
    \sigma_\mathrm{noise}^2 \approx&\left(\frac{2}{\frac{T_\mathrm{m}}{\tau_\mathrm{m}}-1}\right)^{1/2}\\
    &\left(\sigma_\mathrm{spin}^4 + 2\sigma_\mathrm{th}^2\sigma_\mathrm{spin}^2+\left(1+\frac{1}{n_f}\right)\sigma_\mathrm{th}^4\right)^{1/2}.
    \end{split}
\end{equation}
Compared with Eq.~\eqref{eqn:noise}, the coefficient of the last term $\propto\sigma_\mathrm{th}^4$ decreases from $2$ to $\left(1+\frac{1}{n_\mathrm{f}}\right)$, leading to a noise reduction of up to a factor two in the case $n_\mathrm{f} \gg 1$. 

%

\subsection{Compressed Sensing}
\label{sec:compressed}
Both the XYZ and multislice scanning protocols measure a convolution of the sample with smooth and spatially extended PSFs. In Fourier space, this filter-like behavior results in the integrated signal power of the PSF being concentrated in a limited number of Fourier coefficients. As the convolution is a point-wise multiplication in the Fourier domain, the fully sampled measurement data $\mathbf{I}$ is sparse in the Fourier domain as well. Compressed sensing is a signal processing technique~\cite{donoho_compressed_2006,candes_robust_2006} exploiting this sparsity to shorten the acquisition times by recovering the full measurement from a sub-sampled measurement. It is used in various scanning probe and imaging techniques~\cite{andersson_non-raster_2012,oppliger_sparse_2020,ye_compressed_2019}. 


We realize an incoherent sampling pattern by sampling each voxel with probability $p<1$, which produces a binary sampling mask $\mathbf{W}$. Applying this mask to the measurement matrix we obtain the acquired subset $\mathbf{I}_\mathrm{p} = \mathbf{W}\odot \mathbf{I}$, where $\odot$ is the point-wise multiplication operator.\\
To recover non-sampled data points, we search for the measurement $\mathbf{I}$ that is sparsest in the Fourier domain while remaining compatible with the acquired data $\mathbf{{I}}_\mathrm{p}$. Directly minimizing the $\ell_0$ pseudo-norm $\lVert\mathcal{F}\mathbf{I}\rVert_0$ is an infeasible combinatorial task. Instead, the convex $\ell_1$ norm is used, leading to the `basis pursuit denoising' problem~\cite{natarajan_sparse_1995}
\begin{equation}
\label{eq:spare_rec}
\begin{split}
            &\mathbf{{I}^*} = \arg\min_\mathbf{I} \left\lVert\mathcal{F}\mathbf{I}\right\rVert_1\\
        &\text{subject to}\quad\ \lVert  \mathbf{W}\odot \mathbf{I}-\mathbf{{I}_p} \rVert_2^2 \leq \zeta,
\end{split}
\end{equation}
where $\mathcal{F}$ is the 3D discrete Fourier transform and $\zeta$ is a tolerance that accounts for the measurement noise $\bm\eta$.\\
We solve Eq.~\ref{eq:spare_rec} with the SPGL1 software package~\cite{berg_spgl1_2019} which implements a spectral projected-gradient algorithm to solve large-scale $\ell_1$-regularized problems~\cite{van_den_berg_probing_2009}.
In addition to reducing the acquisition time by a factor $\approx1/p$, compressed sensing recovery also denoises the data, due to the measurement uncertainty $\sigma_\mathrm{noise}^2$ mostly being independent between sampling points and therefore not sparse in the Fourier basis. It is consequently suppressed by the $\ell_1$ regularization.

\section{Reconstruction Method}\label{sec:reconstruction}

To recover an image of the scanned sample from the measurement data, one must employ a suitable reconstruction algorithm. In particular, as the scan protocols described by Eqs~\eqref{eq:xyz} and \eqref{eq:multislice} rely on convolutions, a deconvolution step is required to recover the sample $\mathbf{O}_\mathrm{GT}$ from the measurement $\mathbf{I}$. This problem can be formulated as a least squares minimization task,
\begin{equation}
    \tilde{\mathbf{O}} = \arg\min_{\mathbf{O}\geq0} \left\lVert \mathbf{I} -\mathbf{\tilde{I}}\right\rVert_2^2,
    \label{eq:optimization}
\end{equation}
where the solution is the non-negative sample $\tilde{\mathbf{O}}\geq0$ that minimizes the two-norm error $\lVert \mathbf{I}-\mathbf{\tilde{I}} \rVert_2^2 = \sum_{i,j,k} (\mathbf{I}_{i,j,k}-\mathbf{\tilde{I}}_{i,j,k})^2$ with respect to the measurement. Iterative algorithms based on the Landweber iteration~\cite{landweber_iteration_1951} have been applied for the reconstruction of large scale three-dimensional MRFM measurements~\cite{degen_nanoscale_2009,zuger_first_1993}. Although this approach yields acceptable results, large measurement uncertainties lead to slow convergence and large reconstruction errors, cf. Appendix~\ref{app:Landweber}.

An approach to improve the reconstruction is the use of an additional regularization term. Here, we choose the isotropic total variation $\mathrm{TV}(\mathbf{O}) = \sum_{i,j,k} \left\lVert \Delta[O](i,j,k)\right\rVert_2$, where $\Delta[O](i,j,k)$ is the difference vector along each direction, cf. Appendix~\ref{app:admm}. This term penalizes the sum of the Euclidean norm of the samples gradient vectors. It affects the reconstructed sample by smoothing out noise in regions of homogeneous spin density while preserving the sharp edges of the samples features. This is important to avoid overfitting of noisy data. The regularized problem for our multislice protocol is given by 
\begin{equation}
    \tilde{\mathbf{O}} = \arg\min_{\mathbf{O}\geq0} \left[\left\lVert \mathbf{I} -\mathbf{\tilde{I}}\right\rVert_2^2 + \mathrm{TV}(\mathbf{O})\right],
    \label{eq:optimization_ext}
\end{equation}

\begin{figure}[t]
	\centering
	\includegraphics[width=\columnwidth]{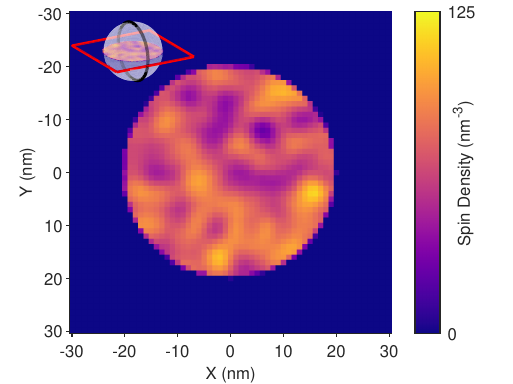}
	\caption{Mid-plane of the spin density ground truth $\mathbf{O}_\mathrm{GT}$ of the spherical test. The object has a diameter of $\SI{40}{\nano\metre}$ and a mean hydrogen spin density of $\SI{60}{\per \nano\metre\cubed}$. The spin density varies with a standard deviation of $\sigma = \SI{18.9}{\per\nano\metre\cubed}$ over a characteristic distance of $\SI{5}{\nano\metre}$. The inset shows the three-dimensional position of the displayed slice in the  sample.}
	\label{Fig:2} 
\end{figure}

To obtain $\tilde{\mathbf{O}}$, we apply the alternating direction method of multipliers (ADMM)~\cite{boyd_distributed_2010,eckstein_douglasrachford_1992,chen_direct_2016}. This approach decomposes the problem in Eq.~\eqref{eq:optimization_ext} into smaller subproblems that individually optimize the different terms of Eq.~\eqref{eq:optimization_ext}. The subproblems are solved sequentially, while ensuring consistency between the subproblems after each step. A detailed description of this algorithm is presented in Appendix~\ref{app:admm}.

Finally, to quantify the reconstruction fidelity, we compute the root-mean square error (RMSE) between the reconstructed volume $\mathbf{\tilde{O}}$ and the ground truth $\mathbf{O}_\mathrm{GT}$
\begin{equation}
    \mathrm{RMSE}(\mathbf{O}_\mathrm{GT},\mathbf{\tilde{O}}) = \sqrt{\frac{1}{N}\sum_{\mathbf{r}}\left(\mathbf{O}_\mathrm{GT}-\mathbf{\tilde{O}}\right)^2},
    \label{eq:error}
\end{equation}
where $N=n_xn_yn_z$ for the XYZ protocol or $N = n_x,n_yn_f$ for the multislice protocol is the total voxel count.

\section{Results}
\label{sec:res}
We evaluate the impact of different setup geometries and scanning protocols by comparing the reconstructed spin distributions of simulated MRFM scans against the known `ground truth', that is, the original spin density in the sample $\mathbf{O}_\mathrm{GT}$, see Fig.~\ref{Fig:2}. Key simulation parameters used are summarized in Table~\ref{tab:1}.

\begin{table}[h]
    \centering
    \begin{tabular}{lcccc}
        \hline
        Parameter & Symbol & Value\\
        \hline
        Scan step size (\si{\nano\metre}) & $\Delta (x,y,z)$ & $1$ \\
        Measurement bandwidth (\si{\hertz}) & $f_\mathrm{meas}$ & $40$ \\
        Sample diameter (\si{\nano\metre}) & $d_\mathrm{O}$ & $40$ \\
        Mean sample spin density (\si{\per\nano\metre\cubed}) & $\overline\rho_\mathrm{spin}$ & $60$ \\
        Sample density variation (\si{\per\nano\metre\cubed}) & $\sigma_\mathrm{\rho}$ & $18.9$\\
        Mechanical force noise density (\si{\atto\newton\per\sqrt\hertz}) & $S_\mathrm{th}$ & $10$ \\
        Spin correlation time (\si{\milli\second}) & $\tau_\mathrm{m}$ & 20\\
        Nanomagnet diameter (\si{\nano\metre}) & $d_\mathrm{mag}$ & 100\\
        Nanomagnet edge radius (\si{\nano\metre}) & $r_\mathrm{edge}$ & 10\\
        Nanomagnet magnetization (\si{\tesla}) & $M_\mathrm{sat}$ & 1.35 \\
        Nanomagnet FEM grid size (\si{\nano\metre}) & $d_\mathrm{FEM}$ & 0.4 \\
        Minimum standoff distance (\si{\nano\metre}) & $z_\mathrm{min}$ & 15\\
        External magnetic field (\si{\tesla}) & $B_0$ &  3\\
        \hline
    \end{tabular}
    \caption{Simulation Parameters}
    \label{tab:1}
\end{table}
For a direct comparison, both cantilever and membrane geometries are assigned the same thermomechanical force noise power spectral density (PSD), $S_\mathrm{th} = \SI{10}{\atto\newton/\sqrt{\hertz}}$ (see Table~\ref{tab:1}). While this is a typical value achieved in experiments utilizing ultrasensitive cantilevers, we note that advanced membrane and string resonators have demonstrated force noise PSDs that are several orders of magnitude lower~\cite{eichler_ultra-high-q_2022}. All simulations maintain a minimum standoff distance $z_\mathrm{min} = \SI{15}{\nano\metre}$ to account for experimental constraints regarding non-contact interactions between the resonator and surface~\cite{krass_force-detected_2022,heritier_spatial_2021}. We note that membrane-based scanning probe setups have recently demonstrated stable operation at separations down to \SI{1}{\nano\metre}~\cite{halg_membrane-based_2021}.

\begin{figure*}[!htb]
	\centering
	\includegraphics[width=\textwidth]{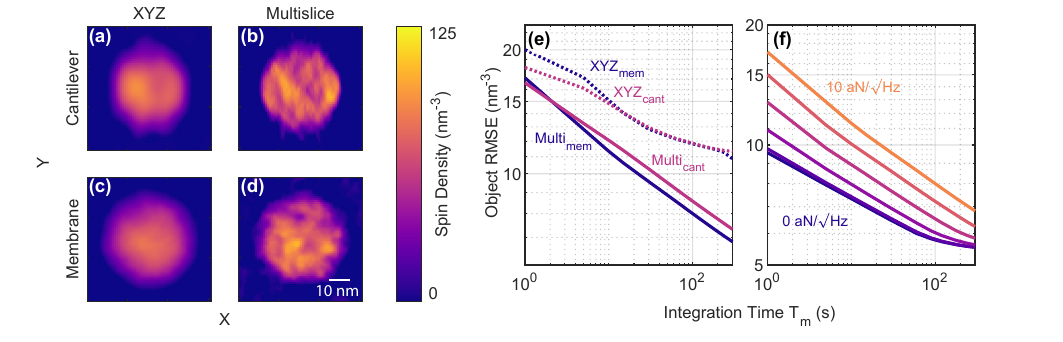}
    \caption{Comparison of scanning protocols. (a-d) Mid-plane $(x,y)$ slices of the reconstructed three-dimensional volumes for an effective integration time of $T_\mathrm{m}=\SI{30}{\second}$ per measurement point (at the reconstruction iteration that produces minimal RMSE error). (a-b) utilize cantilever resonators with the XYZ scanning protocol (a) and the proposed multislice protocol (b). (c-d) employs membrane resonators, again following the XYZ (c) and multislice (d) protocols. In (e) reconstruction results are compared between scanning protocols and measurement geometries for resonators with thermal force noise density of $S_\mathrm{F} =\SI{10}{\atto\newton\per\sqrt\hertz}$ over varying measurement times $T_\mathrm{m}$. (f) shows the reconstruction error using the membrane geometry and the multislice scanning protocol [blue line in (e)] for resonators with decreasing $S_\mathrm{th}$ between $S_\mathrm{th} = \SI{10}{\atto\newton\per\sqrt\hertz}$ and $S_\mathrm{th} = \SI{0}{\atto\newton\per\sqrt\hertz}$, with decrements of $\SI{2}{\atto\newton\per\sqrt\hertz}$.}
	\label{Fig:ProtRes} 
\end{figure*}

The test sample used in this study has a spherical shape, chosen to emulate biologically relevant specimens (e.g. viruses or proteins). The spherical volume is filled with a spatially varying density $\rho_\mathrm{spin}$ of nuclear spins to showcase the performance of the proposed algorithms for non-homogeneous samples. Figure~\ref{Fig:2} illustrates the sample geometry and spin-density distribution. For all simulated scans, the inversion pulse center frequencies $f_{0,l}, ~ l=1,\ldots,n_f$, are tuned to contain at least one resonant slice extending beyond \SI{70}{\nano\metre} from the magnet surface, ensuring complete coverage of the spherical sample. 

\subsection{Resonator Geometries}
Utilizing the pendulum-style cantilever geometry, we benchmark the reconstruction performance of the XYZ and multislice protocols for integration times $T_\mathrm{m}$ ranging from $\SI{1}{\second}$ to $\SI{300}{\second}$ per data point. For the XYZ protocol, we employ a single inversion pulse with center frequency $f_0 = \SI{133.9}{\mega\hertz}$ and bandwidth $f_\mathrm{BW} = \SI{500}{\kilo\hertz}$. The multislice protocol is comprised of $n_\mathrm{f} = 46$ pulses with center frequencies $f_{0,k}$, corresponding to resonance slices extending between $\SI{24}{\nano\metre}$ and $\SI{70}{\nano\metre}$ from the nanomagnet's surface with $\SI{1}{\nano\metre}$ steps and uniform pulse bandwidths of $f_\mathrm{BW} = \SI{500}{\kilo\hertz}$. The used pulse parameters are listed in Appendix \ref{app:IPP}. This results in both protocols acquiring approximately the same number of signal-bearing points.

We apply the ADMM-based reconstruction algorithm with heuristically determined hyper-parameters to the simulated measurements. The chosen hyperparameters and the selection method are presented in Appendix~\ref{app:Landweber}. Figure~\ref{Fig:ProtRes}(a) shows the $(x,y)$ mid-plane of the reconstructed spin distribution from the scan utilizing the XYZ scanning protocol and an integration time of $T_\mathrm{m} = \SI{30}{s}$ per data point. While the sample's spherical shape and average spin density are recovered, its radius is underestimated and nearly all internal density information is lost.

The fully sampled multislice scanning protocol produces the reconstructed spin density shown in Fig.~\ref{Fig:ProtRes}(b). Clearly, this method improves the estimation of the sample's radius and the internal structure of the object. However, compared to the ground truth, small, spatially localized features are still lacking in fidelity, as these are effectively averaged by the PSF. This makes their contributions very small and therefore highly susceptible to measurement noise.

Repeating the same analysis for data simulated with a membrane-style resonator yields the results shown in Fig.~\ref{Fig:ProtRes}(c-d). To match the spatial volume covered by the resonant slices of the cantilever scans, the same center frequencies $f_{0,k}$ as in the cantilever case are used.
Figure~\ref{Fig:ProtRes}(c) shows the mid-plane of the reconstructed spin distribution for the XYZ protocol. Compared to the cantilever geometry, the reconstruction recovers the radius of the spherical shape more accurately, but internal density variations remain largely unresolved. In the multislice reconstruction in Fig.~\ref{Fig:ProtRes}(d), the recovery of the sample's outer shape and average density are strongly improved. Here, the local spin density fluctuations are captured with slightly higher fidelity than in the cantilever case.

Figure~\ref{Fig:ProtRes}(e) quantifies the differences in reconstruction fidelity between scanning protocols and resonator geometries, by plotting the RMSE between reconstruction and ground truth [see Fig.~\ref{Fig:2}] for varying integration times $T_\mathrm{m}$. The XYZ protocol (dashed lines) produces the highest errors throughout for both resonators. Mostly, two factors contribute to this effect. First, the single resonance slice encompasses weaker magnetic field gradients than other protocols, as described in Sec.~\ref{sec:image_protocols}. Second, the single smooth and spatially extended PSF limits sensitivity to intra-volume spin variations. Once the voxel integration time exceeds $T_\mathrm{m}>\SI{60}{\second}$, improvements in the RMSE of the reconstruction plateau, as the regularization term in Eq.~\eqref{eq:optimization_ext} introduces a fixed bias in the solution. By contrast, the multislice scanning protocol (solid lines) consistently achieves a lower RMSE and follows as $\text{RMSE} \propto T_\mathrm{m}^{-0.145\pm0.001}$ for the cantilever and $\text{RMSE} \propto T_\mathrm{m}^{-0.163\pm0.006}$ for the membrane resonators respectively. However, although the RMSE of the multislice protocol decreases slightly faster than that of the XYZ protocol, both decrease slower as a function of $T_\mathrm{m}$ than the measurement uncertainty  $\sigma_\mathrm{noise}^2\propto T_\mathrm{m}^{-0.5}$ of an individual data point [see Eq.~\ref{eq:noisepoint}]. We speculate that the slow decrease in RMSE reflects the ill-conditioning of our measurement problem (see Appendix \ref{app:ill-cond}).

When comparing the reconstruction across scanning protocols in Fig.~\ref{Fig:ProtRes}(e), the membrane and cantilever geometries show similar reconstruction performance for all integration times under the multislice sampling scheme, and achieve comparable bias-limited accuracies in the XYZ protocol at long integration times. The minor gain in reconstruction fidelity of the scans utilizing the multislice technique and membrane resonators indicates a less ill-posed inverse problem than for the cantilever geometry.

A further potential advantage of modern membrane-style resonators is the reduced thermomechanical force-noise PSD $S_\mathrm{th}$, translating into lower reconstruction errors. As $S_\mathrm{th}$ decreases, the thermomechanical noise term $\sigma_\mathrm{th}^2$ in the point-wise noise model [Eq.~\eqref{eq:noisepoint}] becomes negligible and the total measurement uncertainty $\sigma_\mathrm{noise}^2$ becomes increasingly dominated by the standard error in the spin-force variance $\sigma_\mathrm{spin}^2$. Figure~\ref{Fig:ProtRes}(f) illustrates this trend for multislice membrane simulations, with parameters identical to those used in Fig.~\ref{Fig:ProtRes}, except for sweeping $S_\mathrm{th}$ from $S_\mathrm{th} = \SI{10}{\atto\newton\per\sqrt\hertz}$ to $S_\mathrm{th} = \SI{0}{\atto\newton\per\sqrt\hertz}$ in decrements of $\SI{2}{\atto\newton\per\sqrt\hertz}$. The RMSE continues to decrease up to $S_\mathrm{th} = \SI{2}{\atto\newton\per\sqrt\hertz}$, below which the error plateaus, since the total uncertainty $\sigma_\mathrm{noise}$ is dominated by the standard error of the spin-force variance. For this reason, no significant further benefit is obtained even for a hypothetically noiseless resonator (c.f. blue trace) for this spin ensemble size. However, the value of $S_\mathrm{th}$ at which the plateau occurs depends on the scan's voxel size and the spin density of the sample. Imaging smaller or less dense features requires the ability to discern smaller spin force differences $\Delta\sigma_\mathrm{spin}^2$, making smaller $S_\mathrm{th}$ more beneficial. In practice, the uncertainty of a single measurement point in the XYZ scanning protocol $\sigma_\mathrm{noise}^2$ (see Eq.~\eqref{eqn:noise}) is dominated by the standard error of the spin force variance whenever $\sigma_\mathrm{th}^2 < \frac{1}{2}\Delta\sigma_\mathrm{spin}^2(\sqrt{7}-1)$, see Eq.\eqref{eq:SExyz} in Appendix~\ref{app:nosie_est}. 


\begin{figure*}[!ht]
	\centering
	\includegraphics[width=\textwidth]{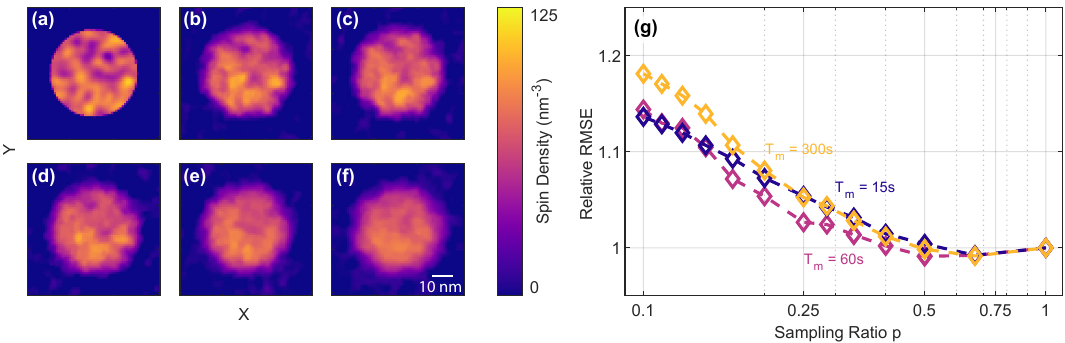}
	\caption{Impact of compressed sensing on the multislice scanning protocol. (a) Mid-plane of the spin density ground truth. (b-f) Mid-plane $(x,y)$ slices of reconstructed volume, obtained by applying compressed sensing to a simulated measurement employing membrane resonators, a data point integration time of $T_\mathrm{m}=\SI{60}{s}$, and sampling ratios (b) $p = 1$, (c) $p= 1/2$, (d) $p= 1/3$, (e) $p= 1/6$, and (f) $p= 1/10$. (g) Relative reconstruction errors of the sub-sampled measurement compared to a scan with identical point integration times $T_\mathrm{m}$ and sampling ratio $p=1$.}
	\label{Fig:CompRes} 
\end{figure*}

\subsection{Compressed Sensing}
Incorporating compressed sensing into the membrane multislice protocol (c.f. Sec.~\ref{sec:compressed}) results in the reconstructions summarized in Fig.~\ref{Fig:CompRes}. We used simulation and reconstruction parameters identical to those in Fig.~\ref{Fig:ProtRes}(d), but adopted a fixed point-integration time $T_\mathrm{m} = \SI{60}{s}$ and sub-sampled the scanning grid with a decreasing sampling ratio $p$. The resulting mid-planes of the reconstructions are presented in Fig.~\ref{Fig:CompRes}(a-d). No significant change is observed between the results recovered from the fully sampled simulation [Fig.~\ref{Fig:CompRes}(b)] and sub-sampling the identical dataset with $p=1/2$ in \ref{Fig:CompRes}(c). However, for lower sampling ratios [$p=1/3$ in (d) and $p=1/6$ in (e)], the fidelity of internal features is visibly degraded.

At a sampling ratio of $p=1/2$ and $T_\mathrm{m} = \SI{60}{s}$ the RMSE between the recovered sample and the ground truth is reduced by $\approx20\%$ relative to the fully sampled case of Fig.~\ref{Fig:ProtRes}(d) using a point integration time of $T_\mathrm{m} = \SI{30}{s}$, despite both experiments sharing the same total acquisition time, $NT_\mathrm{m,eff} = NpT_\mathrm{m} = (N/2)\SI{60}{\second}$.
This improvement likely stems from the longer integration time and therefore higher SNR of individual measurements in the subsampled case and the denoising benefit of the compressed‐sensing technique, which together enhance the fidelity of fine internal features while suppressing noise.

Figure~\ref{Fig:CompRes}(g) quantifies the RMSE as a function of the sampling ratio. For $T_\mathrm{m} = \SI{60}{s}$, the RMSE remains at or below the value of the fully sampled case for $p\geq0.5$ and increases for a further decrease in the number of sampled measurement points. The same trend holds for the integration times per point of $T_\mathrm{m} = \SI{15}{s}$ and $T_\mathrm{m} = \SI{300}{s}$.
Because the compressed sensing algorithm used reconstructs a sparse Fourier representation $\mathbf{I}$ from the subsampled data $\mathbf{I}_\mathrm{p}$, consistent behavior across $T_\mathrm{m}$ indicates that performance is limited by the intrinsic sparsity of the measurement itself. This is determined by the interplay between the ground-truth sample used in these simulations and the point-spread functions employed in the multislice scheme.
 

\section{Outlook}
Overall, membrane-style resonators that oscillate along the $z$-axis have emerged as a promising alternative to pendulum-style cantilever geometries. We have shown that even under identical operating conditions (force noise, standoff distances, and magnetic gradient), membranes deliver equal or better reconstruction fidelity than cantilevers. Further gains should be attainable by exploiting the smaller stand-off distance that membrane resonators can tolerate, which increases the available magnetic field gradients, and by optimizing the magnetic field gradient sources to define more well-conditioned reconstruction problems. In the future, membrane resonators are also expected to yield thermal force noises smaller than that of cantilevers, owing to their excellent quality factors and low optical absorption. Because the success of nuclear spin detected MRFM ultimately depends on reconstructing large-scale three-dimensional data sets with high fidelity, improvements in resonator design must be paired with improved scanning protocols. In this regard, the multislice protocol provides a promising approach. Its acquisition speed can be pushed further by combining it with MRFM phase-multiplexing techniques~\cite{moores_accelerated_2015}, allowing for parallel recording of the response to multiple inversion slices. The achievable speed-up is, however, limited by the rf power required to drive multiple slices in parallel, as increased power results in a tradeoff between the resonators mode temperature and higher parallelization. To improve on this problem, novel approaches to perform MRFM detection can be utilized~\cite{visani_near-resonant_2025,halg_strong_2022}. Taking conservative assumptions, we project a cumulative acceleration factor of 100 combining all these approaches compared to three-dimensional MRFM scanning experiments utilizing the XYZ scanning protocol~\cite{degen_nanoscale_2009}. This speed-up consists of a factors $5$ from using the multislice scanning protocol, a factor $2$ from compressed sensing, a factor $3$ from replacing cantilevers with membrane-type resonators (see Fig.~\ref{Fig:ProtRes}), and a factor $4$ from phase multiplexing. The exact gain will scale with the sample geometry, resonator properties, and the implemented magnetic field gradient source.

Finally, by augmenting the proposed reconstruction algorithm with a sparsity-enforcing regularizer, we could unify reconstruction and compressed sensing into a single optimization problem. This approach is particularly well suited for imaging inherently sparse samples, such as electron-spin–labeled biological structures.

\section*{Acknowledgment} We gratefully acknowledge fruitful discussions with Raffi Budakian, Fabian Natterer, Ender Konukoglu, Raphael Pachlatko, Diego Visani, and Vincent Dumont. This work was supported by the Swiss National Science Foundation (CRSII5\_177198/1 and 200021\_200412/1) and an ETH Zurich Research Grant (ETH-51 19-2).

\bibliography{references.bib}

\clearpage

\appendix

\onecolumngrid

\Large \begin{center} Supplementary information \end{center}\normalsize

 \section{Combined Noise Estimates}\label{app:nosie_est}
\begin{figure}[!htb]
	\centering
	\includegraphics[width=\columnwidth]{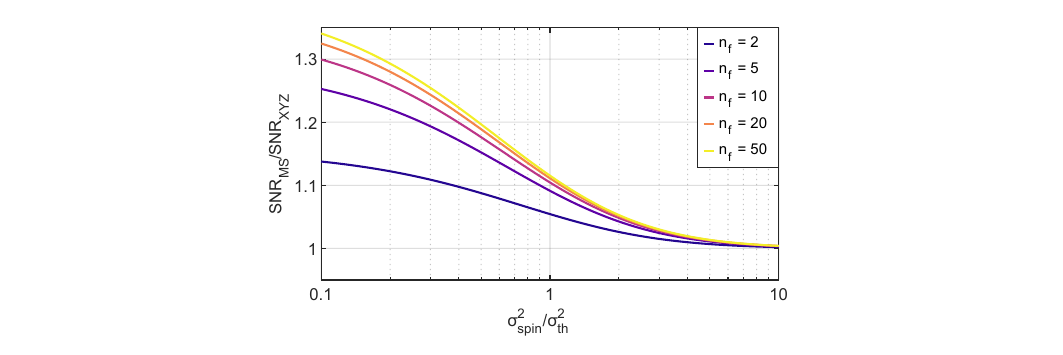}
	\caption{Computed multislice SNR enhancement as a function of the variance ratio $r_\mathrm{var} = \sigma^2_\mathrm{spin}/\sigma^2_\mathrm{th}$ for several values of addressed inversion slice $n_\mathrm{f}$ per resonator position.}
	\label{Fig:AppdxSNR} 
\end{figure}

To determine the number of spins undergoing inversion due to an applied rf-inversion pulse, we record both quadratures of the resonator motion using a lock-in amplifier. After an appropriate phase rotation, the two quadratures correspond to the in phase and out of phase components with respect to the periodic spin-inversions. Their variances are
\begin{equation}
    \sigma_\mathrm{in}^2 = \sigma_\mathrm{spin}^2 + \sigma_\mathrm{th}^2
\end{equation}\begin{equation}
    \sigma_\mathrm{out}^2 = \sigma_\mathrm{th}^2
\end{equation}
Here, $\sigma_\mathrm{spin}^2$ is the force variance caused by the fluctuations of the inverted spin populations, while $\sigma_\mathrm{th}^2$ is the resonators intrinsic thermomechanical force noise.
For the multislice detection scheme described in this paper, $n_\mathrm{f}$ sequential measurements of duration $T_\mathrm{m}$ are performed at each position of the resonator along the $(x,y)$ scanning grid. If $f_\mathrm{BW}>2\Delta f_{0}$ each measurement addresses a spatial distinct inversion slice resulting in $\sigma_\mathrm{spin}^2$ being an independent quantity. Each measurement yields $T_\mathrm{m}/\tau_\mathrm{m}$ statistically independent samples, where $\tau_\mathrm{m}$ is the correlation time of the spin ensemble. Therefore the standard error of the in phase quadrature is described by 
\begin{equation}
    \text{SE}(\sigma_\mathrm{in}^2) = \sqrt{\frac{2}{\frac{T_\mathrm{m}}{\tau_\mathrm{m}}-1}\left( \sigma_\mathrm{spin}^2 + \sigma_\mathrm{th}^2\right)^2}.
    \label{eq:SEin}
\end{equation}
In contrast, the change in the thermomechanical force noise $\sigma_\mathrm{th}^2$ is dominated by non-contact interaction between the resonator and the surface of the magnetic field gradient source and therefore stays constant during the $n_\mathrm{f}$ measurements performed at the same position. Combining the $n_\mathrm{f}$ out of phase estimates thus reduces its standard error to
\begin{equation}
    \text{SE}({\sigma_\mathrm{out}^2}) = \sqrt{\frac{2}{n_\mathrm{f}\frac{T_\mathrm{m}}{\tau_\mathrm{m}}-1}}\sigma_\mathrm{th}^2.
    \label{eq:SEout}
\end{equation}
Under the assumption of independence between $\sigma_\mathrm{spin}^2$ and $\sigma_\mathrm{th}^2$ the unbiased estimator of the spin signal is given by $\sigma_\mathrm{spin}^2 = \sigma_\mathrm{in}^2-\sigma_\mathrm{out}^2$. Propagating the uncertainties in Eqs.~\eqref{eq:SEin}-\eqref{eq:SEout} gives the standard error of the spin signal
\begin{equation}
\begin{split}
    \text{SE}_\mathrm{MS}({\sigma_\mathrm{spin}^2}) &= \sqrt{\left(\text{SE}\left({\sigma_\mathrm{in}^2}\right)^2 +\text{SE}\left({\sigma_\mathrm{out}^2}\right)^2\right)}\\
    &= \sqrt{\frac{2}{\frac{T_\mathrm{m}}{\tau_\mathrm{m}}-1} (\sigma_\mathrm{spin}^4 + \sigma_\mathrm{th}^4+2\sigma_\mathrm{spin}^2\sigma_\mathrm{th}^2) +\frac{2}{\frac{T_\mathrm{m}}{\tau_\mathrm{m}}n_\mathrm{f}-1}\sigma_\mathrm{th}^4}
    \end{split}
\label{eq:SEspin}
\end{equation}
and the resulting signal-to-noise ratio (SNR) of the multislice protocol
\begin{equation}
\begin{split}
    \text{SNR}_\mathrm{MS} = \frac{\sigma_\mathrm{spin}^2}{\text{SE}(\sigma_\mathrm{spin}^2)} &= \left[\frac{2}{\frac{T_\mathrm{m}}{\tau_\mathrm{m}}-1}\left(1+2\frac{\sigma_\mathrm{th}^2}{\sigma_\mathrm{spin}^2} + \frac{\sigma_\mathrm{th}^4}{\sigma_\mathrm{spin}^4}\right) +  \frac{2}{\frac{T_\mathrm{m}}{\tau_\mathrm{m}}n_\mathrm{f}-1}\left(\frac{\sigma_\mathrm{th}^4}{\sigma_\mathrm{spin}^4}\right)\right]^{-1/2}\\
    &\approx\sqrt{\frac{\frac{T_\mathrm{m}}{\tau_\mathrm{m}}-1}{2}}\left(1+2\frac{\sigma_\mathrm{th}^2}{\sigma_\mathrm{spin}^2} + \left(1+\frac{1}{n_\mathrm{f}}\right)\frac{\sigma_\mathrm{th}^4}{\sigma_\mathrm{spin}^4}\right)^{-1/2}
    \end{split}
\end{equation}
For comparison, the standard error and SNR for each measurement point in the XYZ protocol is given by Degen \emph{et al.}~\cite{degen_role_2007}
\begin{equation}
    \text{SE}_\mathrm{XYZ}({\sigma_\mathrm{spin}^2})= \sqrt{\frac{2}{\frac{T_\mathrm{m}}{\tau_\mathrm{m}}-1} (\sigma_\mathrm{spin}^4 + 2\sigma_\mathrm{th}^4+2\sigma_\mathrm{spin}^2\sigma_\mathrm{th}^2)};
    \label{eq:SExyz}
    \end{equation}
    \begin{equation}
    \text{SNR}_\mathrm{XYZ} = \sqrt{\frac{\frac{T_\mathrm{m}}{\tau_\mathrm{m}}-1}{2}}\left(1+2\frac{\sigma_\mathrm{th}^2}{\sigma_\mathrm{spin}^2} + 2\frac{\sigma_\mathrm{th}^4}{\sigma_\mathrm{spin}^4}\right)^{-1/2}.
\end{equation}
Figure~\ref{Fig:AppdxSNR}, plots the ratio of $\text{SNR}_\mathrm{MS}$ to $\text{SNR}_\mathrm{XYZ}$ as a function of $r_\mathrm{var} = \sigma^2_\mathrm{spin}/\sigma^2_\mathrm{th}$ and the number of slices $n_\mathrm{f}$. When the spin fluctuations dominates ($r_\mathrm{var}\gtrsim3$) the improvement is negligible due to the signal uncertainty being limited by the finite sampling of the spin fluctuations. In the thermal noise limited regime, an improvement of up to $\approx 30 \%$ is obtained for realistic $n_\mathrm{f}$ values. Achieving the same gain by extending $T_\mathrm{m}$ instead would require a $\approx 70\%$ longer acquisition time.   

\section{Reconstruction}
\label{app:admm}

\subsection{Problem Definition}
As described in Section \ref{sec:reconstruction}, MRFM scans $\mathbf{I}$ are related to the underlying three-dimensional spin distribution $\mathbf{O}$ via a linear model determined by the measurement protocols (see Eqs.~\ref{eq:xyz}-\ref{eq:multislice}). To recover the spin distribution $\mathbf{O}$ from the measurement requires thereby a reconstruction process. In principle, one could invert these linear operators to recover it, but in practice the problem is severely ill‑posed and highly sensitive to measurement noise.
To obtain a stable estimate of the true spin density, we therefore cast the reconstruction as an optimization problem. Denote by $\mathbf{\tilde{I}}(\mathbf{O})$ the predicted data for a spin distribution $\mathbf{O}$. We seek $\tilde{\mathbf{O}}$ that minimizes the squared discrepancy between measured and predicted data.
\begin{equation}
    \tilde{\mathbf{O}} = \arg\min_{\mathbf{O}} \left\lVert \mathbf{I} -\mathbf{\tilde{I}}(\mathbf{O})\right\rVert_2^2,
    \label{eq:optimization_apdx}
\end{equation}
Although this least-squares problem enforces a close match between measurement and predicted data, noise in the measurement will result in non-optimal solutions which may contain non-negative or oscillating components that contain no physical meaning. To remedy these problems we regularize the problem by adding two additional penalty terms. 

First, the solution is constrained to be non-negative $\mathbf{O}\geq0$ as the solution corresponds to a density. We enforce this by introducing the indicator function on the non-negative half-space to the problem
\begin{equation}
    \Pi_+(\mathbf{A}) = 
      \begin{cases}
     0 &  \mathbf{A}\geq 0\\
    \infty & \text{else}.
  \end{cases}
\end{equation}
Second, noise and modal mismatch tend to produce high frequency artifacts. To suppress these while preserving sharp edges in the sample, we penalize the sum of the Euclidean norm of the discrete gradient at each voxel
\begin{equation}
\text{TV}(\mathbf{A})_{2,1} = \sum_{i,j,k} \sqrt{ \left( A_{i+1,j,k} - A_{i,j,k} \right)^2 + \left( A_{i,j+1,k} - A_{i,j,k} \right)^2 + \left( A_{i,j,k+1} - A_{i,j,k} \right)^2 }.
\label{eq:tv}
\end{equation}
This regularization term is called the isotropic total variation. Combining these terms yields the final optimization problem
\begin{equation}
    \tilde{\mathbf{O}} = \arg\min_{\mathbf{O}} \left[\left\lVert \mathbf{I} -\mathbf{\tilde{I}}\right\rVert_2^2 + \Pi_+(\mathbf{O)} + \text{TV}(\mathbf{O)}_{2,1}\right],
    \label{eq:optimization_ext_apdx}
\end{equation}
This formulation balances matching measurement and predicted data with physical and regularity constraints, making the inversion more robust to noise and model imperfections.

\subsection{Vectorization}
To simplify the derivation of reconstruction algorithms, it is convenient to transform the three dimensional arrays into tall vectors. Therefore, for both $\mathbf{I}$ and $\mathbf{O}$ the vectorized versions $\mathbf{i}$ and $\mathbf{o}$ are defined by stacking all entries in lexicographic order (fastest in $z$, then $y$ and $x$). In particular, for an arbitrary three dimensional array $\mathbf{X}$ of size $(n_x,n_y,n_z)$, we form  
\begin{equation}
    \mathbf{x} = \begin{bmatrix}
{X}(1,1,1) &{X}(1,1,2)   \cdots  {X}(1,1,n_z) & {X}(2,1,1)  \cdots {X}(n_x,1,n_z) &  {X}(1,2,1)  \cdots  {X}(n_x,n_y,n_z)
\end{bmatrix}^\mathsf{T}.
\label{eq:vect}
\end{equation}
This produces a column vector $\mathbf{x}$ of length $N=(n_xn_yn_z)$. In the multislice protocol, where the third dimension corresponds to the inversion pulse parameter set index, the same stacking results in a vector of length $N' = n_xn_yn_f$.
Using this vectorization, the Equations~\eqref{eq:xyz}-\eqref{eq:multislice} corresponding to the scanning protocols can be written in vectorized form
\begin{equation}
    \mathbf{i} = \mathbf{A}\mathbf{o} + \boldsymbol{\eta},
\end{equation}
where $\boldsymbol\eta$ is the vectorized noise term, and $\mathbf{A}$ is the measurement protocol matrix encoding the measurement protocol. For the XYZ scan we define $\mathbf{A} = \mathbf{A}_\mathrm{XYZ}$ a $N\times N$ matrix, while for the multislice protocol it is given by $\mathbf{A} = \mathbf{A}_\mathrm{MS}$ of size $N'\times N$. In the following, we derive both these process matrices.

\subsubsection{XYZ Scan}
The conventional XYZ scan corresponds to a three-dimensional convolution of the sample's spin distribution $\mathbf{O}$ and the PSF $\mathbf{H}$ of a single inversion slice. Therefore the problem can be written as
\begin{equation}
    \mathbf{I} =  \mathbf{H}\star_{(x,y,z)}\mathbf{O}
\end{equation}
where $\star_{x,y,z}$ is the three dimensional circular convolution operator under the assumption of sufficient padding to minimize wrapping errors.

Applying the convolution theorem, the equation can be transformed into a point-wise multiplication in the Fourier domain

\begin{equation}
    \mathbf{I} = \mathcal{F}^{-1}\left\{\mathbf{\hat{H}}\odot\mathbf{\hat{O}}\right\}
\end{equation}
Here, $\mathcal{F}$ is the three dimensional Fourier transform, $\mathbf{\hat{O}} = \mathcal{F}\mathbf{O}$ is the Fourier domain representation, and $\odot$ the element-wise (Hadamard) product.\\
As the point-wise multiplication of two matrices is equivalent to multiplication of the vectorized matrix with a diagonal matrix we can write the vectorized XYZ scan as
\begin{equation}
    \mathbf{i} = \mathbf{F}^{-1}\textrm{diag}\hat{\left(\mathbf{h}\right)}\mathbf{F}\mathbf{{o}}
\end{equation}
where $\mathbf{F}$ is the 3D DFT matrix, $\mathbf{F}^{-1}$ its inverse, and $\mathbf{\hat{h}}$ the vectorized form of $\mathbf{\hat{H}}$. Hence, the measurement protocol matrix for the XYZ scan is 
\begin{equation}
\mathbf{A}_\mathrm{XYZ} = \mathbf{F}^{-1}\textrm{diag}\left(\mathbf{\hat{h}}\right)\mathbf{F}.
\end{equation}

\subsubsection{Multislice Scan}
\label{sec:MS_prot_mat}
In the multislice protocol, instead of physically scanning in the $z$ direction,  at every position $(x,y)$ the response to different set of inversion pulse parameters is measured. For the $l$-th pulse the PSF is $\mathbf{H}^l$ and the signal at position $(x,y)$ is the sum over all elements of the point-wise multiplication of the PSF and the sample's spin distribution. This results in the protocol being described by the sum of all 2D convolutions over every $(x,y)$ plane
\begin{equation}
\mathbf{I}_{\mathrm{MS},l} = \sum_z \mathbf{H}^l\star_{(x,y)}\mathbf{O}, 
\end{equation}
where $\star_{(x,y)}$ denotes the circular convolution in slice-wise order between the $(x,y)$ planes.

Applying the Fourier transform $\mathcal{F}_{(x,y)}$ and the convolution theorem in the $x$ and $y$ dimension rewrites the protocol into
\begin{equation}
    \mathbf{I}_{\mathrm{MS},i}= \mathcal{F}^{-1}_{(k_x,k_y)}\left\{\sum_z\left(\mathbf{\hat{H}}^l(k_x,k_y,z)\odot\mathbf{\hat{O}}(k_x,k_y,z)\right)\right\}.
\end{equation}
Here, $\mathbf{\hat{H}}^l(k_x,k_y,z) = \mathcal{F}_{(x,y)}\mathbf{{H}}_i(x,y,z)$ and $\mathbf{\hat{O}}(k_x,k_y,z) = \mathcal{F}_{(x,y)}\mathbf{{O}}(x,y,z)$ are the PSF and sample matrices where each $(x,y)$-plane is Fourier transformed.\\
To express this equation as a matrix vector equation $\mathbf{i} = \mathbf{A_\mathrm{MS}o}$, $\mathbf{\hat{H}}^l(k_x,k_y,z)\odot\mathbf{\hat{O}}(k_x,k_y,z)$ can be transformed to a multiplication with a diagonal matrix $\textrm{diag}\left(\mathbf{\hat{H}}^l(k_x,k_y,z)\right)\mathbf{\hat{o}}$
\begin{equation}
\begin{split}
    \mathbf{i} &= \mathbf{F}^{\mathsf{H}}_{(x,y)}\sum_z\textrm{diag}\left(\mathbf{\hat{H}}^l(k_x,k_y,z)\right)\mathbf{\hat{o}}\\
    &= \mathbf{F}^{\mathsf{H}}_{(x,y)}\sum_z\textrm{diag}\left(\mathbf{\hat{H}}^l(k_x,k_y,z)\right)\mathbf{F}_{(x,y)}\mathbf{o}\\
    &= \mathbf{F}^{\mathsf{H}}_{(x,y)}\mathbf{G}\mathbf{F}_{(x,y)}\mathbf{o}.\\
    \end{split}
\end{equation}
Here, $\mathbf{F}_{(x,y)}$ is the 2D-DFT matrix applying the two-dimensional Fourier transform to each $(x,y)$ plane of a vector representing a three-dimensional data set.
The matrix $\mathbf{G} = \sum_z\textrm{diag}\left(\mathbf{\hat{H}}^l(k_x,k_y,z)\right)$ can be written as a block-diagonal operator
\begin{equation}
        \mathbf{G} = 
    \begin{bmatrix}
        \mathbf{G}_{(1,1)} & 0   & \cdots & 0 \\
        0   & \mathbf{G}_{(1,2)} & \cdots & 0 \\
        \vdots & \vdots & \ddots & \vdots \\
        0   & 0   & \cdots & \mathbf{G}_{(n_x,n_y)}
    \end{bmatrix}, 
    \end{equation}
    where each block $\mathbf{G_{(i,j)}}$ of size $(n_f\times n_z)$ has entries
    \begin{equation}
    \mathbf{G}_{(i,j)} =
    \begin{bmatrix}
        \hat{H}^1(i,j,1) & \hat{H}^1(i,j,2)   & \cdots & \hat{H}^1(i,j,n_z) \\
        \hat{H}^2(i,j,1)   & \hat{H}^2(i,j,2) & \cdots & \hat{H}^2(i,j,n_z) \\
        \vdots & \vdots & \ddots & \vdots \\
        \hat{H}^{n_f}(i,j,1)   & \hat{H}^{n_f}(i,j,2)   & \cdots & \hat{H}^{n_f}(i,j,n_z)
    \end{bmatrix} 
\end{equation}
 The resulting measurement process matrix of the multislice scanning protocol is given by $\mathbf{A}_\mathrm{MS} = \mathbf{F}^\mathsf{H}_{(x,y)}\mathbf{G}\mathbf{F}_{(x,y)}$.

\subsection{Reconstruction}
The optimization problem~\ref{eq:optimization_ext_apdx} consists of three distinct terms: a least-squares data-fidelity term, a non-negativity constraint, and an isotropic total-variation penalty. Solving the whole problem in a single step is infeasible. Instead, the alternating direction method of multipliers (ADMM) \cite{boyd_distributed_2010,eckstein_douglasrachford_1992} is applied. ADMM splits the problem into simple sub-problems that are solved sequentially while enforcing convergence to a global optimum. 

To recast Eq.~\ref{eq:optimization_ext_apdx} in ADMM form, we introduce two auxiliary variables. To ensure the non-negativity of the solution $\mathbf{o}_+$ is introduced, while $\mathbf{o}_\mathrm{TV}$ is added to describe the local gradients required to compute the total-variation penalty term. Furthermore, the two constraints $\mathbf{\mathbf{o}_\mathrm{TV} = \mathbf{Do}}$ and $\mathbf{o}_+ = \mathbf{o}$ are added to ensure the additional auxiliary variables coincide with the spin density estimate $\mathbf{o}$. With these definitions, the constrained optimization problem is given as
\begin{equation}
\begin{split}
    \arg\min_{\mathbf{o}} &\frac{1}{2}\left\lVert\mathbf{A}_\mathrm{MS}\mathbf{o} -\mathbf{\tilde{i}}\right\rVert_2^2 + \gamma_\mathrm{TV} \lVert \mathbf{\mathbf{o}_\mathrm{TV}} \rVert_{2,1} + \Pi_+(\mathbf{o}_+),\\
    s.t.\quad&\mathbf{o}_+ - \mathbf{o} = \mathbf{0}\\
    &\mathbf{\mathbf{o}_\mathrm{TV}} - \mathbf{Do} = \mathbf{0},\\
    \end{split}
    \label{eq:optimization_admm}
\end{equation}
Here, we replace the notation of the total variation operator $\text{TV}(\mathbf{o})_{2,1}$ with the matrix vector product of the vector $\mathbf{o}$ with the difference matrix $\mathbf{D} = (\mathbf{D}_x,\mathbf{D}_y,\mathbf{D}_z)^\mathsf{T}$. This matrix computes the finite difference along the $x$,$y$, and $z$ directions, respectively. The group norm of this term $\lVert \mathbf{Do} \rVert_{2,1} = \sum_i \lVert (\mathbf{Do})_i \rVert_2= \sum_{i,j,k} \sqrt{ \left( o_{i+1,j,k} - o_{i,j,k} \right)^2 + \left( o_{i,j+1,k} - o_{i,j,k} \right)^2 + \left( o_{i,j,k+1} - o_{i,j,k} \right)^2 } $, first applies the two norm along the added difference dimension, before computing the sum for every point in the measurement.

To solve this problem while incorporating the constraints, Eq.~\ref{eq:optimization_admm} is transformed into the unconstrained augmented Lagrangian notation~\cite{nocedal_numerical_2006}.    
\begin{equation}
\begin{split}
    L_{\boldsymbol{\nu}}(\mathbf{o},\mathbf{o}_+,\mathbf{o}_\mathrm{TV},\boldsymbol{\lambda}_1,\boldsymbol{\lambda}_2) &= \frac{1}{2}\left\lVert \mathbf{A}_\mathrm{MS}\mathbf{o} -\mathbf{i}\right\rVert_2^2 + \gamma_\mathrm{TV} \lVert \mathbf{o}_\mathrm{TV} \rVert_{2,1} + \Pi_+(\mathbf{o}_+)\\ 
    &+ \boldsymbol{\lambda}_1^\mathsf{T}(\mathbf{o}_+-\mathbf{o}) + \frac{\nu_1}{2} \left\lVert \mathbf{o}_+-\mathbf{o} \right\rVert_2^2\\
    &+ \boldsymbol{\lambda}_2^\mathsf{T}(\mathbf{o}_\mathrm{TV}-\mathbf{Do}) +\frac{\nu_2}{2} \left\lVert \mathbf{o}_\mathrm{TV}-\mathbf{Do} \right\rVert_2^2,\\
\end{split}
\end{equation}
where $\boldsymbol{\lambda}_1,\boldsymbol{\lambda}_2$ are the Lagrange multipliers and $\boldsymbol{\nu} = (\nu_1,\nu_2)$ are positive penalty parameters.
The optimal solution to~\ref{eq:optimization_admm} corresponds to solving the saddle point problem $(\mathbf{o}^*,\mathbf{o}_+^*,\mathbf{o}_\mathrm{TV}^*,\boldsymbol{\lambda}_1^*,\boldsymbol{\lambda}_2^*) = \arg\min_{\mathbf{o},\mathbf{o}_\mathrm{TV},\mathbf{o}_+} \arg\max_{\boldsymbol{\lambda}_1,\boldsymbol{\lambda}_2} L_{\boldsymbol{\nu}}(\mathbf{o},\mathbf{o}_+,\mathbf{o}_\mathrm{TV},\boldsymbol{\lambda}_1,\boldsymbol{\lambda}_2)$~\cite{boyd_distributed_2010}.

The ADMM algorithm solves this problem in an iterative manner. In each iteration $(k)$ we successively minimize $L_{\boldsymbol{\nu}}$ with respect to $\mathbf{o},\mathbf{o}_{+}$ and $\mathbf{o}_{\mathrm{TV}}$, before performing a standard gradient ascent step on the dual variables $\boldsymbol{\lambda}_1,\boldsymbol{\lambda}_2$. This procedure is summarized in Algorithm~\ref{alg:ADMM}.

\begin{algorithm}
\caption{ADMM Reconstruction}\label{alg:ADMM}
\begin{algorithmic}
\State $\mathbf{o}^0,\mathbf{o}_+^0,\mathbf{o}_\mathrm{TV}^0,\boldsymbol{\lambda}_1^0,\boldsymbol{\lambda}_2^0\gets \mathbf{0}$
\Repeat
    \State $\mathbf{o}^{(k+1)} \gets \underset{\mathbf{o}}{\arg \min} \, L_{\boldsymbol{\nu}}(\mathbf{o},\mathbf{o}_+^{(k)},\mathbf{o}_\mathrm{TV}^{(k)},\boldsymbol{\lambda}_1^{(k)},\boldsymbol{\lambda}_2^{(k)})$
    \State $\mathbf{o}_+^{(k+1)} \gets \underset{\mathbf{o}_+}{\arg \min} \, L_{\boldsymbol{\nu}}(\mathbf{o}^{(k+1)},\mathbf{o}_+,\mathbf{o}_{\mathrm{TV}}^{(k)},\boldsymbol{\lambda}_1^{(k)},\boldsymbol{\lambda}_2^{(k)})$
    \State $\mathbf{o}_{\mathrm{TV}}^{(k+1)} \gets \underset{\mathbf{o}_\mathrm{TV}}{\arg \min} \, L_{\boldsymbol{\nu}}(\mathbf{o}^{(k+1)},\mathbf{o}_+^{(k+1)},\mathbf{o}_\mathrm{TV},\boldsymbol{\lambda}_1^{(k)},\boldsymbol{\lambda}_2^{(k)})$
    \State $\boldsymbol{\lambda}_1^{(k+1)}  \gets \boldsymbol{\lambda}_1^{(k)}+\nu_1 \left(\mathbf{o}_+^{(k+1)} -\mathbf{o}^{(k+1)}\right)$
    \State $\boldsymbol{\lambda}_2^{(k+1)}  \gets \boldsymbol{\lambda}_2^{(k)}+\nu_2 \left(\mathbf{o}_\mathrm{TV}^{(k+1)} -\mathbf{Do}^{(k+1)}\right)$
\Until{$i_{err}\leq err_{max}$}
\end{algorithmic}
\end{algorithm}
As this algorithm requires the solution of an optimization problem for each sub-problem update it is necessary to find an explicit update equation for all subproblems. While using the ADMM to solve problems of similar form to the XYZ problem is a known problem~\cite{antipa_diffusercam_2018}, the multislice problem requires the derivation of the individual update steps. 

\subsubsection{$\mathbf{o}$-update}
The variable $\mathbf{o}$ is the main quantity of interest. It holds the current estimate of the three-dimensional spin-density distribution. In iteration $(k)$ we obtain the next iteration $\mathbf{o}^{(k+1)}$ by minimization of the augmented Lagrangian. This optimization problem is given by 
\begin{equation}
\begin{split}
    \underset{\mathbf{o}}{\arg \min} \,& L_{\boldsymbol{\nu}}(\mathbf{o},\mathbf{o}_+^{(k)},\mathbf{o}_\mathrm{TV}^{(k)},\boldsymbol{\lambda}_1^{(k)},\boldsymbol{\lambda}_2^{(k)})\\
    = \underset{\mathbf{o}}{\arg \min} \,& \frac{1}{2}\left\lVert \mathbf{A}_\mathrm{MS}\mathbf{o} -\mathbf{i}\right\rVert_2^2
    - \left(\boldsymbol{\lambda}_1^{(k)}\right)^{\mathsf{T}}\mathbf{o} + \frac{\nu_1}{2} \left\lVert \mathbf{o}_+^{(k)}-\mathbf{o} \right\rVert_2^2
    - \left(\boldsymbol{\lambda}_1^{(k)}\right)^{\mathsf{T}}\mathbf{D}\mathbf{o} +\frac{\nu_2}{2} \left\lVert \mathbf{o}_\mathrm{TV}^{(k)}-\mathbf{Do} \right\rVert_2^2.
\end{split}
\end{equation}
Because all terms are at most quadratic in $\mathbf{o}$, setting the derivative to zero yields the update equation
\begin{equation}
\label{eq:o-update}
    \mathbf{o}^{(k+1)} = \left( \mathbf{A}_\mathrm{MS}^{\mathsf{H}}\mathbf{A}_\mathrm{MS} + \nu_1 + \nu_2\mathbf{D}^{\mathsf{H}}\mathbf{D}\right)^{-1}\left( \mathbf{A}_\mathrm{MS}^{\mathsf{H}}\mathbf{A}_\mathrm{MS}\mathbf{i} + \nu_1 + \boldsymbol{\lambda}_1^{(k)}+ \mathbf{D}^{\mathsf{T}}\left(\nu_2 +\boldsymbol{\lambda}_2^{(k)}\right)\right).
\end{equation}

Directly applying the measurement process matrix $\mathbf{A}_\mathrm{MS}$ is infeasible for a moderately sized problem due to its size of $(N'\times N)$. As shown in Sect.~\ref{sec:MS_prot_mat} it can be factorized as $\mathbf{A}_\mathrm{MS} = \mathbf{F}_{(x,y)}^{-1}\mathbf{G}\mathbf{F}_{(x,y)}$, using the block-diagonal matrix $\mathbf{G}$. To compute $\mathbf{A}_\mathrm{MS}^{\mathsf{H}}\mathbf{A}_\mathrm{MS}$, this factorization can be exploited in connection with the anti-distributivity of the hermitian adjoint resulting in

\begin{equation}
    \mathbf{A}_\mathrm{MS}^{\mathsf{H}}\mathbf{A}_\mathrm{MS} =  (\mathbf{F}^{\mathsf{H}}_{(x,y)}\mathbf{G}\mathbf{F}_{(x,y)})^{\mathsf{H}}\mathbf{F}^{\mathsf{H}}_{(x,y)}\mathbf{G}\mathbf{F}_{(x,y)} = \mathbf{F}_{(x,y)}^{\mathsf{H}} \mathbf{G}^{\mathsf{H}} \mathbf{F}_{(x,y)}\mathbf{F}^{\mathsf{H}}_{(x,y)}\mathbf{G}\mathbf{F}_{(x,y)} = \mathbf{F}^{\mathsf{H}}_{(x,y)}\mathbf{G}^{\mathsf{H}}\mathbf{G}\mathbf{F}_{(x,y)}, 
\end{equation}
where the adjoint of $\mathbf{G}$ is given by
\begin{equation}
    \mathbf{G}^{\mathsf{H}} = 
    \begin{bmatrix}
        \mathbf{G}_{(1,1)}^{\mathsf{H}} & 0   & \cdots & 0 \\
        0   & \mathbf{G}_{(1,2)}^{\mathsf{H}} & \cdots & 0 \\
        \vdots & \vdots & \ddots & \vdots \\
        0   & 0   & \cdots & \mathbf{G}_{(n_x,n_y)}^{\mathsf{H}}
    \end{bmatrix}, \mathbf{G}_{(i,j)}^{\mathsf{H}} =
    \begin{bmatrix}
        \hat{H}^{1*}(i,j,1) & \hat{H}^{2*}(i,j,1)   & \cdots & \hat{H}^{n_f*}(i,j,1) \\
        \hat{H}^{1*}(i,j,1)   & \hat{H}^{2*}(i,j,2) & \cdots & \hat{H}^{n_f*}(i,j,2) \\
        \vdots & \vdots & \ddots & \vdots \\
        \hat{H}^{1*}(i,j,n_z)   & \hat{H}^{2*}(i,j,n_z)   & \cdots & \hat{H}^{n_f*}(i,j,n_z)
    \end{bmatrix} \in \mathbb{C}^{n_z\times n_f}
\end{equation}
Hence $\mathbf{A}_\mathrm{MS}^{\mathsf{H}}\mathbf{A}_\mathrm{MS}$ is block-diagonalized by the same 2D-DFT matrix $\mathbf{F}_{(x,y)}$. This reduces the computational complexity of the forward model to the computation of $n_x n_y$ dense multiplications of vectors with matrices of size $n_f \times n_f$.

Equation~\ref{eq:o-update} contains the inverse $\left( \mathbf{A}_\mathrm{MS}^{\mathsf{H}}\mathbf{A}_\mathrm{MS} + \nu_1\bm{\mathcal{I}} + \nu_2\mathbf{D}^{\mathsf{H}}\mathbf{D}\right)^{-1}$. The approach to finding a traceable method to compute this inverse is to show that all three terms are (block-) diagonalized by the 2D Fourier transform. We already showed this property for $(\mathbf{A}_\mathrm{MS}^{\mathsf{H}}\mathbf{A}_\mathrm{MS})$.

For the second term $\nu_1 = \nu_1\bm{\mathcal{I}}$, where $\bm{\mathcal{I}}$ is the identity matrix of size $n_xn_yn_f \times n_xn_yn_f$, the diagonalization by $\mathbf{F}_{(x,y)}$ is given by the unitary property of the 2D-DFT matrix $\nu_1\bm{\mathcal{I}} = \nu_1\mathbf{F}_{(x,y)}^{-1}\bm{\mathcal{I}}\mathbf{F}_{(x,y)}$.\\
The third term $\nu_2\mathbf{D}^{\mathsf{H}}\mathbf{D}$ computes the two dimensional difference for each $z$-level. We can use the fact that $\mathbf{D}$ and $\mathbf{D}^{\mathsf{T}}$ are block-circulant matrices and diagonalizable by the two-dimensional Fourier transform~\cite{yang_fast_2009}. Due to the vectorization pattern \ref{eq:vect}, the finite difference matrix used in the multislice algorithm is a block diagonal matrix where each block is the two-dimensional difference matrix. Hence $\mathbf{D}^{\mathsf{T}}\mathbf{D} = \mathbf{F}_{(x,y)}^{\mathsf{H}}\mathbf{\tilde{D}}\mathbf{F}_{(x,y)}$ and we can compute the inverse as

\begin{equation}
    \left( \mathbf{A}_\mathrm{MS}^{\mathsf{H}}\mathbf{A}_\mathrm{MS} + \nu_1\bm{\mathcal{I}} + \nu_2\mathbf{D}^{\mathsf{H}}\mathbf{D}\right)^{-1} = \left( \mathbf{F}^{\mathsf{H}}_{(x,y)}\left(\mathbf{G}^{\mathsf{H}} \mathbf{G} +  \nu_1\bm{\mathcal{I}} + \nu_2\mathbf{\tilde{D}} \right)\mathbf{F}_{(x,y)} \right)^{-1} =  \mathbf{F}^{\mathsf{H}}_{(x,y)}\left(\mathbf{G}^{\mathsf{H}} \mathbf{G} +  \nu_1\bm{\mathcal{I}} + \nu_2\mathbf{\tilde{D}} \right)^{-1}\mathbf{F}_{(x,y)}
\end{equation}
This reduces the problem to computing the inverse of a block diagonal matrix in the Fourier domain.  
\begin{equation}
\begin{split}
    \left(\mathbf{G}^{\mathsf{H}} \mathbf{G} +  \nu_1\bm{\mathcal{I}} + \nu_2\mathbf{\tilde{D}} \right)^{-1}& = 
    \begin{bmatrix}
        \left(\mathbf{G}_{(1,1)}^{\mathsf{H}}\mathbf{G}_{(1,1)} + \nu_1\bm{\mathcal{I}} + \nu_2\mathbf{\tilde{D}}_{(1)}\right)^{-1}    & \cdots & 0 \\
        \vdots  & \ddots & \vdots \\
        0      & \cdots & \left(\mathbf{G}_{(n_x,n_y)}^{\mathsf{H}}\mathbf{G}_{n_x,n_y)} + \nu_1\bm{\mathcal{I}} + \nu_2\mathbf{\tilde{D}}_{(n_xn_y)}\right)^{-1}
    \end{bmatrix},
    \end{split}
\end{equation}
where each block $\left(\mathbf{G}_{(i,j)}^{\mathsf{H}}\mathbf{G}_{(i,j)} + \nu_1\bm{\mathcal{I}} + \nu_2\mathbf{\tilde{D}}_{(ij)}\right)^{-1}$ is a dense matrix of size $(n_f \times n_f)$. Thus, the inverse computed in the Fourier domain consists of $n_xn_y$ inverses of size $n_f\times n_f$ each. This makes its computation feasible for large-scale problems. In addition, this computation depends only on the $n_f$ PSFs of the applied inversion pulses. Since these remain constant during the duration of the scan, it is sufficient to compute the inverse once at the beginning of the reconstruction algorithm, instead of at each iteration $k$.
In summary, the full update equation for the $\mathbf{o}$-update is given by
\begin{equation}
\begin{split}
    \mathbf{o}_{(k+1)} &= \left( \mathbf{A}_\mathrm{MS}^{\mathsf{H}}\mathbf{A}_\mathrm{MS} + \nu_1 + \nu_2\mathbf{D}^{\mathsf{H}}\mathbf{D}\right)^{-1}\left( \mathbf{A}_\mathrm{MS}^{\mathsf{H}}\mathbf{A}_\mathrm{MS}\mathbf{i} + \nu_1 + \boldsymbol{\lambda}_1^{(k)}+ \mathbf{D}^{\mathsf{T}}\left(\nu_2 +\boldsymbol{\lambda}_2^{(k)}\right)\right)\\
    &= \mathbf{F}_{(x,y)}^{-1}(\mathbf{G}^{\mathsf{H}}\mathbf{G} + \nu_1 + \nu_2\mathbf{\tilde{D}})^{-1}\mathbf{F}_{(x,y)}(\mathbf{F}_{(x,y)}^{-1}\mathbf{G}^{\mathsf{H}}\mathbf{G}\mathbf{F}_{(x,y)}\mathbf{i} + \nu_1 + \boldsymbol{\lambda}_1^{(k)}+ \mathbf{D}^{\mathsf{T}}\left(\nu_2 +\boldsymbol{\lambda}_2^{(k)}\right))
    \end{split}
\end{equation}

\subsubsection{$\mathbf{o}_+$-Update}
To perform the $\mathbf{o}_+$-update step, the augmented Lagrangian is minimized with respect to $\mathbf{o}_+$, taking into account the already updated spin distribution $\mathbf{o}^{(k+1)}$. 
\begin{equation}
\begin{split}
    &\underset{\mathbf{o}_+}{\arg \min} ~ L_{\boldsymbol{\nu}}(\mathbf{o}^{(k+1)},\mathbf{o}_+,\mathbf{o}_\mathrm{TV}^{(k)},\boldsymbol{\lambda}_1^{(k)},\boldsymbol{\lambda}_2^{(k)})\\
    &= \underset{\mathbf{o}_+}{\arg \min} ~ \Pi_+(\mathbf{o}_+)
    + {\left(\boldsymbol{\lambda}_1^{(k)}\right)}^{\mathsf{T}}(\mathbf{o}_+) + \frac{\nu_1}{2} \left\lVert \mathbf{o}_+-\mathbf{o}^{(k+1)} \right\rVert_2^2\\
    &=\underset{\mathbf{o}_+}{\arg \min} ~ \Pi_+(\mathbf{o}_+) + \frac{\nu_1}{2} \left\lVert \mathbf{o}_+-\left(\mathbf{o}^{(k+1)}+\frac{\boldsymbol{\lambda}_1^{(k)}}{\nu_1} \right)\right\rVert_2^2\\ 
\end{split}
\end{equation}
The solution to this problem of form $\arg \min_x f(x) + 1/2\lVert x-y\rVert_2^2$ is the proximal operator of $f$ evaluated at $y$. While in general this problem requires solving an additional optimization problem, for many functions $f(x)$ there exists a closed form solution~\cite{parikh_proximal_2014}. For the indicator function of the non-negative half-space $\Pi_+(\mathbf{o}_+)$ the solution is given by
\begin{equation}
    \mathbf{o}_+^* = \max\left(0,\mathbf{o}^{(k+1)}+\frac{\boldsymbol{\lambda}_1^{(k)}}{\nu_1}\right).
    \label{eq:p_upd}
\end{equation}
As this is an element-wise operation, it bears no significant computational cost, reducing the impact of the $\mathbf{o}_+$-update on the algorithms runtime.
\subsubsection{$\mathbf{o}_\mathrm{TV}$-update}
The auxiliary variable $\mathbf{o}_\mathrm{TV}$ stores the finite-difference vectors needed for the isotropic TV penalty. For every element in $\mathbf{o}$ it contains two entries, the gradient in $x$ and $y$. The update for $\mathbf{o}_\mathrm{TV}^{(k+1)}$ is obtained by solving
\begin{equation}
\begin{split}
   &\underset{\mathbf{o}_\mathrm{TV}}{\arg \min} ~ L_{\boldsymbol{\nu}}(\mathbf{o}^{(k+1)},\mathbf{o}_+^{(k+1)},\mathbf{o}_\mathrm{TV},\boldsymbol{\lambda}_1^{(k)},\boldsymbol{\lambda}_2^{(k)})\\
    &= \underset{\mathbf{o}_\mathrm{TV}}{\arg \min} ~ \gamma_\mathrm{TV}\lVert \mathbf{o}_\mathrm{TV}\rVert_{2,1}
    + \left(\boldsymbol{\lambda}_2^{(k)}\right)^{\mathsf{T}}\mathbf{Do} + \frac{\nu_1}{2} \left\lVert \mathbf{o}_\mathrm{TV}-\mathbf{Do}^{(k+1)} \right\rVert_2^2\\
    &= \underset{\mathbf{o}_\mathrm{TV}}{\arg \min} ~ \gamma_\mathrm{TV}\lVert \mathbf{o}_\mathrm{TV}\rVert_{2,1}
    +  \frac{\nu_1}{2} \left\lVert \mathbf{o}_\mathrm{TV}-\left(\mathbf{Do}^{(k+1)}+\frac{\left(\boldsymbol{\lambda}_2^{(k)}\right)^{\mathsf{T}}}{\nu_2} \right)\right\rVert_2^2\\
\end{split}
\end{equation}
This again is a proximal step for the mixed $(2,1)$-norm. The proximal operator for $f(\mathbf{o}_\mathrm{TV}) = \lVert \mathbf{o}_\mathrm{TV}\rVert_{2,1}$ is~\cite{parikh_proximal_2014}
\begin{equation}
    \mathbf{o}_\mathrm{TV}^* = \max\left(0,1-\frac{\gamma_\mathrm{TV}}{\nu_2\lVert\mathbf{Do}^{(k+1)}+\boldsymbol{\lambda}_2^{(k)}/\nu_2\rVert_2}\right)(\mathbf{Do}^{(k+1)}+\boldsymbol{\lambda}_2^{(k)}/\nu_2)
\end{equation}
Combining all update steps and inserting them into Alg.~\ref{alg:ADMM} results in the final ADMM algorithm shown in Alg.~\ref{alg:ADMM_final} used to solve the reconstruction problem of the multislice scanning protocol.
\begin{algorithm}
\caption{ADMM Multislice Reconstruction}\label{alg:ADMM_final}
\begin{algorithmic}
\State $\mathbf{o}^{(0)},\mathbf{o}_+^{(0)},\mathbf{o}_\mathrm{TV}^{(0)},\boldsymbol{\lambda}_1^{(0)},\boldsymbol{\lambda}_2^{(0)}\gets 0$
\Repeat
    \State $\mathbf{o}^{(k+1)} \gets \left( \mathbf{A}_\mathrm{MS}^{\mathsf{H}}\mathbf{A}_\mathrm{MS} + \nu_1 + \nu_2\mathbf{D}^{\mathsf{H}}\mathbf{D}\right)^{-1}\left( \mathbf{A}_\mathrm{MS}^{\mathsf{H}}\mathbf{A}_\mathrm{MS}\mathbf{i} + \nu_1 + \boldsymbol{\lambda}_1^{(k)}+ \mathbf{D}^{\mathsf{T}}\left(\nu_2 +\boldsymbol{\lambda}_2^{(k)}\right)\right)$
    \State $\mathbf{o}_+^{(k+1)} \gets \max\left(0,\mathbf{o}^{(k+1)}+\frac{\boldsymbol{\lambda}_1^{(k)}}{\nu_1}\right)$
    \State $\mathbf{o}_\mathrm{TV}^{(k+1)} \gets \max\left(0,1-\frac{\gamma_\mathrm{TV}}{\nu_2\lVert\mathbf{Do}^{(k+1)}+\boldsymbol{\lambda}_2^{(k)}/\nu_2\rVert_2}\right)(\mathbf{Do}^{(k+1)}+\boldsymbol{\lambda}_2^{(k)}/\nu_2)$
    \State $\boldsymbol{\lambda}_1^{(k+1)}  \gets \boldsymbol{\lambda}_1^{(k)}+\nu_1 (\mathbf{o}_+^{(k+1)} -\mathbf{o}^{(k+1)})$
    \State $\boldsymbol{\lambda}_2^{(k+1)}  \gets \boldsymbol{\lambda}_2^{(k)}+\nu_2 (\mathbf{o}_\mathrm{TV}^{(k+1)} -\mathbf{Do}^{(k+1)})$
\Until{$i_\mathrm{err}\geq err_\mathrm{max}$}
\end{algorithmic}
\end{algorithm}

\section{Ill-Conditioned Problem}
\label{app:ill-cond}
An inverse problem is ill-conditioned if its condition number $\kappa = \sigma_\mathrm{max}/\sigma_\mathrm{min}\gg 1$ is very large, where $\sigma_\mathrm{min}$ and $\sigma_\mathrm{max}$ are the extreme singular values of the problem. Applied to the MRFM reconstruction problem, a perturbation of the estimated sample, $\Delta\tilde{\mathbf{O}}$, alters the scan data $\Delta\mathbf{I}$ within the bounds $\sigma_\mathrm{min}\lVert\Delta\tilde{\mathbf{O}}\rVert_2 \leq \lVert \Delta \mathbf{I}\rVert_2 \leq \sigma_\mathrm{max}\lVert\Delta\tilde{\mathbf{O}}\rVert_2$. Updates of the estimate $\tilde{\mathbf{O}}$ corresponding to small singular values produce only minor changes in the predicted data and are therefore highly susceptible to measurement noise, whereas directions aligned with large singular values are more robust. 
In our measurement geometries, the PSFs are smooth and spatially extended. As a result, the overall sample's shape and its average spin density map to large singular values and strongly influence $\mathbf{I}$, while localized spin density variations are averaged and projected onto small singular values, resulting in small changes to the scan. Iterative reconstruction algorithms applied to this problem exhibit semi-convergent behavior: large-scale features in the sample $\mathbf{O}_\mathrm{GT}$ are recovered in few iterations, while localized variations emerge only after many iterations, during which measurement noise is progressively amplified. Beyond an optimal iteration, this noise growth outweighs further gains. This optimal stopping iteration minimizes the RMSE of the reconstruction. Although ADMM's regularization terms suppress parts of this noise amplification, it cannot overcome the ill-conditioning of the problem, and the empirical RMSE decay rate remains below the ideal SNR improvement factor.

\section{ADMM Performance}
\label{app:Landweber}
To evaluate the performance of the ADMM algorithm used in this work, it is compared to the Landweber algorithm~\cite{landweber_iteration_1951}, a special case of gradient descent to solve ill-posed inverse problems. In previous works it was utilized to solve the MRFM reconstruction problem~\cite{degen_nanoscale_2009,mamin_isotope-selective_2009}. Additionally, to encompass the non-negativity of the solution it is extended to the projected Landweber iteration, projecting the solution after every iteration onto the positive half-space. One major downside of this algorithm is the low convergence speed with increasing problem size. To accelerate the convergence we use the Barzilai-Borwein method~\cite{barzilai_two-point_1988}, to dynamically update the step size at each iteration. The complete algorithm is described in Alg.~\ref{alg:Landweber}
\begin{algorithm}
\caption{Adaptive Projected Landweber Iteration}
\label{alg:Landweber}
\begin{algorithmic}
\State $\mathbf{o}^{(0)}, \gamma_\mathrm{LW}^{(0)},i,\gamma_\mathrm{LW,max},\gamma_\mathrm{LW,min}$
\Repeat
    \State $\mathbf{o}^{(k+1)} = \max(\mathbf{o}^{(k)}- \gamma_\mathrm{LW}^{(k)} \mathbf{A}^{\mathsf{H}}(\mathbf{A}\mathbf{o}^{(k)}-\mathbf{i}) ,0)$
    \State $\mathbf{s}^{(k+1)} = \mathbf{o}^{(k+1)}-\mathbf{o}^{(k)}$
    \State $\mathbf{y}^{(k+1)} = \mathbf{A}^{\mathsf{H}}(\mathbf{A} \mathbf{o}^{(k+1)}-\mathbf{i})$
    \State $\gamma_\mathrm{LW}^{(k+1)} = \frac{{\mathbf{s}^{(k+1)}}^{\mathsf{T}}\mathbf{s}^{(k+1)}}{{\mathbf{s}^{(k+1)}}^{\mathsf{T}}\mathbf{y}^{(k+1)}}$
    \State $\gamma_\mathrm{LW}^{(k+1)} = \max(\min(\gamma_\mathrm{LW}^{(k+1)},\gamma_\mathrm{LW,max}),\gamma_\mathrm{LW,min})$
\Until{$i_\mathrm{err}\geq err_\mathrm{max}$}
\end{algorithmic}
\end{algorithm}

\begin{figure}[!htb]
	\centering
	\includegraphics[width=\columnwidth]{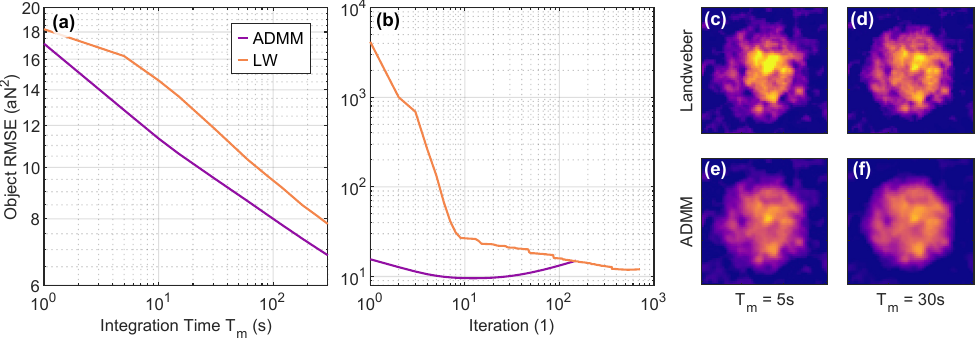}
	\caption{ADMM performance evaluation. (a) reconstruction error for extended adaptive Landweber algorithm and ADMM using the identical simulated measurement. (b) RMSE convergence for both algorithms, when reconstructing scan with $T_m = 30s$. (c)-(f) Reconstructed $(x,y)$ mid-plane of same object used as in main text. (c)-(d) applies the Landweber algorithm, (e)-(f) the ADMM reconstruction framework.}
	\label{Fig:ADMMvsLandweber} 
\end{figure}

Figure~\ref{Fig:ADMMvsLandweber} compares the performance between the ADMM algorithm and the adapted Landweber iteration. In Fig.~\ref{Fig:ADMMvsLandweber}(a), the reconstruction error is compared for a simulated multislice scan using a membrane resonator and parameters identical to the results in~\ref{Fig:ProtRes}. Due to the finer control of the regularization in the ADMM algorithm, the RMSE can be more precisely adapted to the measurement environment, resulting in a reduced RMSE over the full range of integration times $T_\mathrm{m}$. Furthermore, comparing the convergence speed of the two algorithms for the exemplary case of $T_\mathrm{m} = \SI{30}{\second}$ in Fig.~\ref{Fig:ADMMvsLandweber}(b), the ADMM achieves its minimal RMSE after 10 iterations, while the Landweber algorithm reaches its minimum after 510 iterations. On a workstation running an Intel Core i9-14900k and 128 GB RAM, the ADMM algorithm achieves its minimal error after $\SI{67}{\second}$, while the Landweber algorithm takes $\SI{779}{\second}$.\\

\begin{figure}[!htb]
	\centering
	\includegraphics[width=\columnwidth]{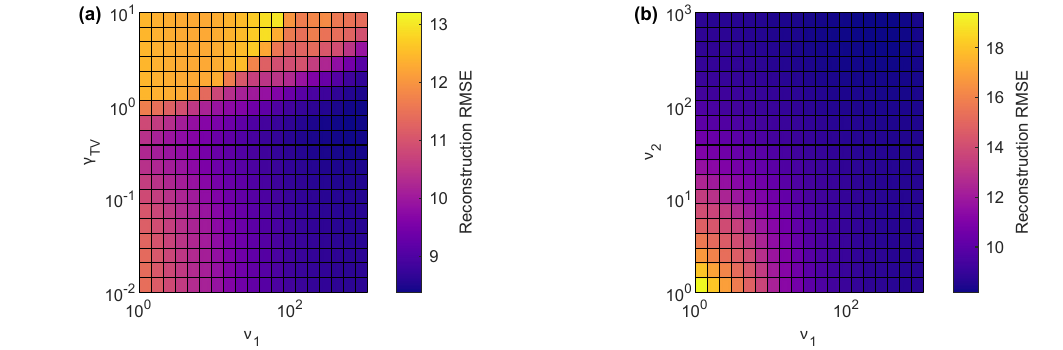}
	\caption{Sensitivity analysis for hyperparameters used for the ADMM multislice reconstruction of the membrane based dataset. (a) The RMSE of the reconstruction depending on the hyperparameters $\nu_1$ and $\gamma_\mathrm{TV}$. $\nu_2$ is kept fixed at $\nu_2 = 25$ for each parameter combination. (b) Sensitivity analysis for $\nu_1$ and $\nu_2$, while $\gamma_\mathrm{TV}$ is fixed at a value of $\gamma_\mathrm{TV} = 0.05$.}
	\label{Fig:SensAna} 
\end{figure}
It is important to find a set of hyperparameters in the vicinity of the optimal point. The size of the area around the optimal point where hyperparameters lead to near-optimal results defines the tolerance of the method. Fortunately, this area in which a useful result is obtained spans a large part of the parameter space, making it possible to find an adequate set of hyperparameters. To find the optimal set, techniques such as Bayesian hyperparameter optimization could be used. Due to the three- or four-dimensional parameter space and the slow evaluation per parameter set, this is a time-consuming endeavor. Additionally, the resulting hyperplane is highly non-convex, requiring large amounts of data points to find the global optimum. Instead, in this work, we decided to determine the hyperparameters using heuristic techniques. From a starting point, the parameters are individually changed, and the derivative is estimated and followed. This procedure is repeated for the other parameters until a local optimum is found. While this method does not necessarily provide us with a global optimum, a non-optimal set of hyperparameters typically only results in a minor increase of the RMSE.\\
To showcase the resilience against non-optimal hyperparameters, we performed a sensitivity analysis in the case of the membrane-based study by performing a grid evaluation of hyperparameter combinations. A full grid search over all combinations of the three hyperparameters would be computationally infeasible. Instead, we perform two grid searches, where one parameter is kept fixed while the other two parameters are swept over 20 logarithmically-spaced values each. In the first sensitivity analysis shown in Fig.~\ref{Fig:SensAna}(a), we fix $\nu_2$ to the value used in this study. Here, we observe that while for some parameter combinations with $\gamma_\mathrm{TV}\geq 1$ and $\nu_1 \leq 10$ the RMSE of the reconstruction is significantly increased, there exists a wide range of hyperparameter combinations that result in RMSE close to the optimum. When repeating the same analysis but keeping $\gamma_\mathrm{TV}$ fixed at the value used in this study [see Fig.~\ref{Fig:SensAna}(b)], we can confirm this stability of the multislice algorithm to hyperparameter variations.\\
Due to the observed resilience against non-optimal parameters, we decided to utilize the same set of hyperparameters when changing experimental parameters (e.g. measurement time). The heuristically determined hyperparameters used in this study are presented in Table~\ref{tab:hyperparams}

\begin{table}
    \centering
    \begin{tabular}{||c|c|c|c|c|c||}
    \hline
    Case & $\gamma_\mathrm{TV}$ & $\nu_1$ & $\nu_2$ & $\nu_3$\\
    \hline
         XYZ ADMM Cantilever& 0.015 & 10 & 1 & 1\\
         Multislice ADMM Cantilever&  0.05& 250 &  25& -\\
         XYZ ADMM Membrane&0.015& 10& 1& 1\\
         Multislice ADMM Membrane& 0.05 & 250&25& -\\
     \hline
    \end{tabular}
    \caption{Hyperparameters of the ADMM algorithms used in this work.}
    \label{tab:hyperparams}
\end{table}

When comparing the recovered objects from both methods in Fig.~\ref{Fig:ADMMvsLandweber}(c)-(f), the missing regularization term is observed in the augmented Landweber algorithm. While the Landweber algorithm concentrates the highest spin density information in the center of the reconstructed object, the ADMM algorithm distributes the spin density over the full sample. Additionally, for short per-voxel integration times, the result obtained from the Landweber algorithm does not capture the clear boundary of the object, instead showing a decaying spin density towards the sample's edge (see Fig.~\ref{Fig:ADMMvsLandweber}(c)). In comparison, the ADMM based reconstruction captures the overall shape of the imaged object already for a short integration time of \SI{5}{s} (see Fig.~\ref{Fig:ADMMvsLandweber}(d)).

\section{Inversion Pulse Parameters}
\label{app:IPP}
The simulations in this work that utilize the multislice protocol use $n_f = 46$ different sets of parameters for the applied rf inversion pulses. The parameters of these pulses were chosen to ensure that the furthest point above the magnet surface that has a magnetic field that corresponds to a Larmor frequency equal to the pulse's center frequency $\frac{\gamma }{2\pi}|B(\mathbf{r})|= f_0$ is positioned between \SI{24}{nm} and $\SI{70}{nm}$ above the nanomagnet surface. In addition, the bandwidth of all inversion pulses is uniformly set to $f_\mathrm{BW} = \SI{500}{kHz}$.\\
Table~\ref{tab:PulseFreqs} lists the center frequencies of all pulses used in the multislice protocol and the corresponding extension of the PSF from the magnet's surface.
\begin{table}[!h]
    \centering
    \begin{tabular}{||c|c||c|c||c|c||c|c||c|c||}
    \hline
    PSF Ext.& $f_0$ &PSF Ext.& $f_0$ &PSF Ext.& $f_0$ &PSF Ext.& $f_0$ &PSF Ext.& $f_0$ \\
    \hline
         \SI{24}{nm}&\SI{146.74}{MHz} &\SI{34}{nm}&\SI{142.39}{MHz}&\SI{44}{nm}&\SI{139.05}{MHz}&\SI{54}{nm}&\SI{136.05}{MHz}&\SI{64}{nm}&\SI{134.91}{MHz} \\
         \SI{25}{nm}&\SI{146.25}{MHz} &\SI{35}{nm}&\SI{142.02}{MHz}&\SI{45}{nm}&\SI{138.78}{MHz}&\SI{55}{nm}&\SI{136.58}{MHz}&\SI{65}{nm}&\SI{134.74}{MHz} \\
         \SI{26}{nm}&\SI{145.78}{MHz} &\SI{36}{nm}&\SI{141.66}{MHz}&\SI{46}{nm}&\SI{138.52}{MHz}&\SI{56}{nm}&\SI{136.36}{MHz}&\SI{66}{nm}&\SI{134.59}{MHz} \\
         \SI{27}{nm}&\SI{145.32}{MHz} &\SI{37}{nm}&\SI{141.30}{MHz}&\SI{47}{nm}&\SI{138.25}{MHz}&\SI{57}{nm}&\SI{136.16}{MHz}&\SI{67}{nm}&\SI{134.44}{MHz} \\
         \SI{28}{nm}&\SI{144.86}{MHz} &\SI{38}{nm}&\SI{140.95}{MHz}&\SI{48}{nm}&\SI{137.98}{MHz}&\SI{58}{nm}&\SI{135.98}{MHz}&\SI{68}{nm}&\SI{134.29}{MHz} \\
         \SI{29}{nm}&\SI{144.43}{MHz} &\SI{39}{nm}&\SI{140.60}{MHz}&\SI{49}{nm}&\SI{137.73}{MHz}&\SI{59}{nm}&\SI{135.79}{MHz}&\SI{69}{nm}&\SI{134.15}{MHz} \\
         \SI{30}{nm}&\SI{144.01}{MHz} &\SI{40}{nm}&\SI{140.27}{MHz}&\SI{50}{nm}&\SI{137.48}{MHz}&\SI{60}{nm}&\SI{135.60}{MHz}&\SI{70}{nm}&\SI{133.87}{MHz} \\
         \SI{31}{nm}&\SI{143.59}{MHz} &\SI{41}{nm}&\SI{139.95}{MHz}&\SI{51}{nm}&\SI{137.25}{MHz}&\SI{61}{nm}&\SI{135.42}{MHz}&& \\
         \SI{32}{nm}&\SI{143.18}{MHz} &\SI{42}{nm}&\SI{139.64}{MHz}&\SI{52}{nm}&\SI{137.02}{MHz}&\SI{62}{nm}&\SI{135.25}{MHz}&& \\
         \SI{33}{nm}&\SI{142.77}{MHz} &\SI{43}{nm}&\SI{139.34}{MHz}&\SI{53}{nm}&\SI{136.79}{MHz}&\SI{63}{nm}&\SI{135.07}{MHz}&& \\
         \hline
    \end{tabular}
    \caption{Center frequencies of inversion pulses used in the simulation of MRFM measurements utilizing the multislice protocol.}
    \label{tab:PulseFreqs}
\end{table}

\section{Additional Sample}
\label{app:addSample}
\begin{figure}[!htb]
	\centering
	\includegraphics[width=\columnwidth]{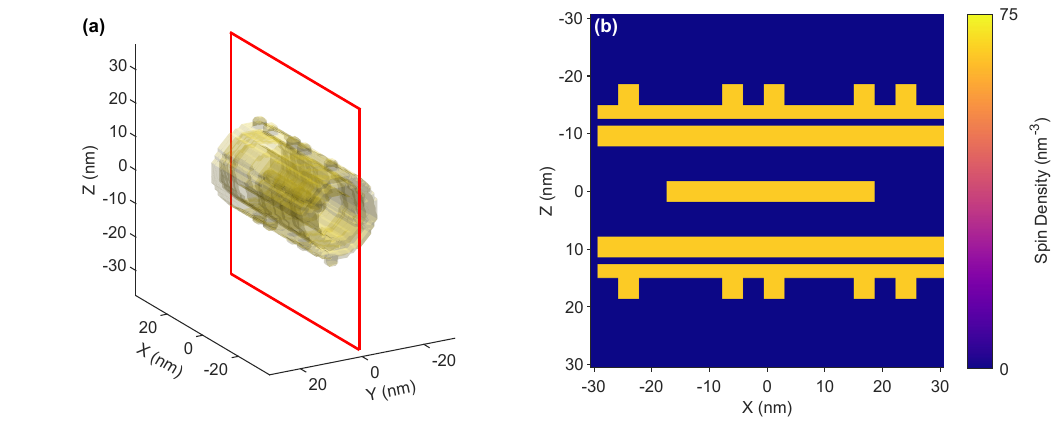}
	\caption{Alternative sample used to showcase performance of the multislice scanning protocol. (a) Three-dimensional view of the entire TMV-inspired sample. It consists out of two concentric cylinders with a rod in the center and extensions at its bottom and top. (b) Central (x,z) plane of the sample, used to visualize algorithm performance. The plane's position is indicated by the red square in (a).}
	\label{Fig:ObjAlt} 
\end{figure}

To demonstrate the applicability of the multislice protocol to different sample geometries, we present reconstruction results for an additional object modeled after a segment of a tobacco-mosaic virus (TMV), as used in previous 3D-MRFM experiments~\cite{degen_nanoscale_2009}, see Fig.~\ref{Fig:ObjAlt}. It has the same average spin density as the spherical sample used in the main text. In this additional simulation study, we used the same magnet geometry and properties, experimental parameters, and inversion pulses as in the previous study. The reconstruction hyperparameters are listed in Table~\ref{tab:hyp2}. 

\begin{table}
    \centering
    \begin{tabular}{||c|c|c|c|c|c||}
    \hline
    Case & $\gamma_\mathrm{TV}$ & $\nu_1$ & $\nu_2$ & $\nu_3$\\
    \hline
         Multislice ADMM Membrane& 1 & 50&10& -\\
            XYZ ADMM Membrane&1& 50& 10& 1\\
     \hline
    \end{tabular}
    \caption{Hyperparameters of the ADMM algorithms used to reconstruct the TMV dataset.}
    \label{tab:hyp2}
\end{table}

\begin{figure}[!htb]
	\centering
	\includegraphics[width=\columnwidth]{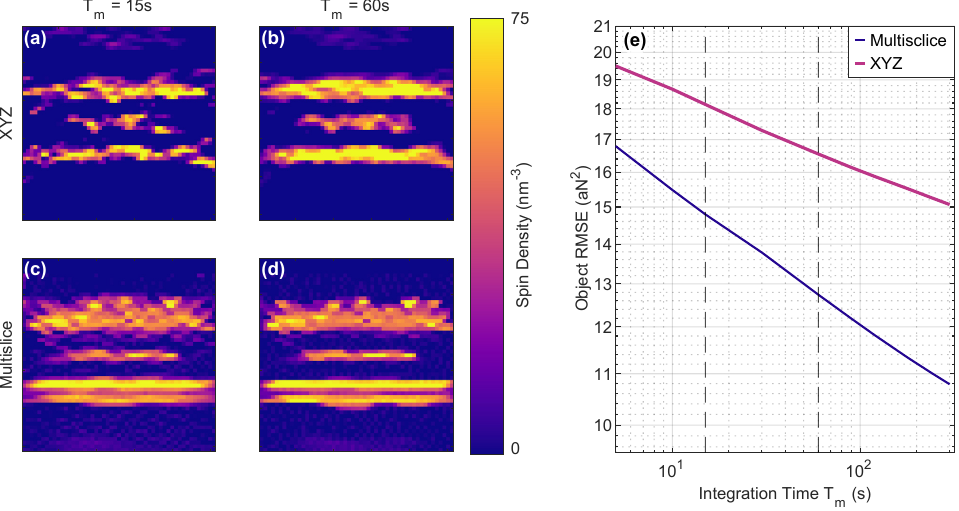}
	\caption{Reconstruction quality for alternative sample. (a-b) Central (x,z) planes of the samples reconstructed from simulated scans using the XYZ scanning protocol for voxel integration times of $T_m = \SI{15}{s}$(a) and $T_m = \SI{60}{s}$ (b). (c-d) The same central (x,z) planes when utilizing the multislice scanning protocol and identical integration times. (e) reconstruction error for both scanning protocols for a wide range of voxel integration times $T_m$. The integration times visualized in (a-d) are indicated by the vertical dashed lines.}
	\label{Fig:AltErr} 
\end{figure}

The results of this study are shown in Fig.~\ref{Fig:AltErr}. For the new sample, Fig.~\ref{Fig:AltErr}(e) shows that the RMSE of the multislice reconstruction is significantly smaller than that of the XYZ protocol over the full range of integration times studied. In addition, we observe a faster decrease in the reconstruction error with increasing voxel integration time, similar to the case of the spherical sample. Figure~\ref{Fig:AltErr}(a-d) show the central (x,z) plane of the reconstructed samples for integration times of $T_m = \SI{15}{s}$ and $T_M = \SI{60}{s}$. Comparing them with the ground truth in Fig.~\ref{Fig:ObjAlt}(b), we observe that the multislice protocol [Fig.~\ref{Fig:AltErr}(c-d)] resolves significantly more detail. In the lower half of the sample, the two cylinder walls can be distinguished and the uniform density is recovered. Furthermore, the central rod is recovered, and the overall position of the upper part and the top extensions are reconstructed. We assume the inability to recover the gap between the top half of the two cylinders stems from the increased distance of this part of the sample from the magnet surface. As a result, the magnetic field gradients are reduced, leading to a thicker signal-generating resonance slice and therefore to decreased spatial resolution. The reconstructed samples obtained from the XYZ scanning protocol show significantly less detail for both voxel integration times shown in Fig.~\ref{Fig:AltErr}(a-b). For $T_m = \SI{15}{s}$ the signal distribution only indicates the rough location of the sample. The two separate cylinder walls are not resolved, nor are the correct wall thicknesses or any of the top or bottom extensions. For $T_m = \SI{60}{s}$, the overall shape estimation is slightly more precise, but there are still no visible details.

\end{document}